
%
\def\unredoffs{}
\tolerance=1000\hfuzz=2pt
\catcode`\@=11 
\ifx\hyperdef\UNd@FiNeD\def\hyperdef#1#2#3#4{#4}\def\hyperref#1#2#3#4{#4}\def\href#1#2{#2}\fi
\magnification=1200\unredoffs\baselineskip=16pt plus 2pt minus 1pt
\def\Date#1{\vfill\leftline{#1}\tenpoint\supereject%
\footline={\hss\tenrm\hyperdef\hypernoname{page}\folio\folio\hss}}%

{\count255=\time\divide\count255 by 60 \xdef\hourmin{\number\count255}
 \multiply\count255 by-60\advance\count255 by\time
 \xdef\hourmin{\hourmin:\ifnum\count255<10 0\fi\the\count255}
}
\def\date{\number\day.\number\month.\number\year\ at \hourmin}


\def\nolabels{\def\wrlabeL##1{}\def\eqlabeL##1{}\def\reflabeL##1{}}
\def\writelabels{\def\wrlabeL##1{\leavevmode\vadjust{\rlap{\smash%
{\line{{\escapechar=` \hfill\rlap{\sevenrm\hskip.03in\string##1}}}}}}}%
\def\eqlabeL##1{{\escapechar-1\rlap{\sevenrm\hskip.05in\string##1}}}%
\def\reflabeL##1{\noexpand\llap{\noexpand\sevenrm\string\string\string##1}}}
\nolabels

\global\newcount\secno \global\secno=0
\global\newcount\meqno \global\meqno=1
\def\s@csym{}

\def\newsec#1\par{\global\advance\secno by1%
{\toks0{#1}\message{(\the\secno. \the\toks0)}}%
\global\subsecno=0\eqnres@t\let\s@csym\secsym\xdef\secn@m{\the\secno}\noindent
{\bf\hyperdef\hypernoname{section}{\the\secno}{\the\secno.} #1}%
\writetoca{{\string\hyperref{}{section}{\the\secno}{\bf \the\secno\quad}} {\bf #1}}\par%
\nobreak\medskip\nobreak\noindent\ignorespaces}
\def\eqnres@t{\xdef\secsym{\the\secno.}\global\meqno=1\bigbreak\bigskip}
\def\sequentialequations{\def\eqnres@t{\bigbreak}}\xdef\secsym{}

\global\newcount\subsecno \global\subsecno=0
\def\subsec#1\par{\global\advance\subsecno by1%
{\toks0{#1}\message{(\s@csym\the\subsecno. \the\toks0)}}%
\global\subsubsecno=0%
\ifnum\lastpenalty>9000\else\bigbreak\fi
\noindent{\it\hyperdef\hypernoname{subsection}{\secn@m.\the\subsecno}%
{\secn@m.\the\subsecno.} #1}\writetoca{\string\hskip1.45cm
{\string\hyperref{}{subsection}{\secn@m.\the\subsecno}{\secn@m.\the\subsecno.}}
{#1}}\par\nobreak\medskip\nobreak\noindent\ignorespaces}

\def\appendix#1#2{\global\meqno=1\global\subsecno=0\xdef\secsym{\hbox{#1.}}%
\bigbreak\bigskip\noindent{\bf Appendix \hyperdef\hypernoname{appendix}{#1}%
{#1.} #2}{\toks0{(#1. #2)}\message{\the\toks0}}%
\xdef\s@csym{#1.}\xdef\secn@m{#1}%
\writetoca{{\string\hyperref{}{appendix}{#1}{\bf {#1}\quad}} {\bf #2}}%
\par\nobreak\medskip\nobreak}

%
\def\checkm@de#1#2{\ifmmode{\def\f@rst##1{##1}\hyperdef\hypernoname{equation}%
{#1}{#2}}\else\hyperref{}{equation}{#1}{#2}\fi}
\def\eqnn#1{\DefWarn#1\xdef #1{(\noexpand\relax\noexpand\checkm@de%
{\s@csym\the\meqno}{\secsym\the\meqno})}%
\wrlabeL#1\writedef{#1\leftbracket#1}\global\advance\meqno by1}
\def\f@rst#1{\c@t#1a\em@ark}\def\c@t#1#2\em@ark{#1}
\def\eqna#1{\DefWarn#1\wrlabeL{#1$\{\}$}%
\xdef #1##1{(\noexpand\relax\noexpand\checkm@de%
{\s@csym\the\meqno\noexpand\f@rst{##1}1}{\hbox{$\secsym\the\meqno##1$}})}
\writedef{#1\numbersign1\leftbracket#1{\numbersign1}}\global\advance\meqno by1}
\def\eqn#1#2{\DefWarn#1%
\xdef #1{(\noexpand\hyperref{}{equation}{\s@csym\the\meqno}%
{\secsym\the\meqno})}$$#2\eqno(\hyperdef\hypernoname{equation}%
{\s@csym\the\meqno}{\secsym\the\meqno})\eqlabeL#1$$%
\writedef{#1\leftbracket#1}\global\advance\meqno by1}
\def\xeqn{\expandafter\xe@n}\def\xe@n(#1){#1}
\def\xeqna#1{\expandafter\xe@n#1}
\def\eqns#1{(\e@ns #1{\hbox{}})}
\def\e@ns#1{\ifx\UNd@FiNeD#1\message{eqnlabel \string#1 is undefined.}%
\xdef#1{(?.?)}\fi{\let\hyperref=\relax\xdef\next{#1}}%
\ifx\next\em@rk\def\next{}\else%
\ifx\next#1\xeqn#1\else\def\n@xt{#1}\ifx\n@xt\next#1\else\xeqna#1\fi
\fi\let\next=\e@ns\fi\next}

\def\DefWarn#1{\ifx\UNd@FiNeD#1\else
\immediate\write16{*** WARNING: the label \string#1 is already defined ***}\fi}
%
\newskip\footskip\footskip14pt plus 1pt minus 1pt 
\def\footnotefont{\ninepoint}\def\f@t#1{\footnotefont #1\@foot}
\def\f@@t{\baselineskip\footskip\bgroup\footnotefont\aftergroup\@foot\let\next}
\setbox\strutbox=\hbox{\vrule height9.5pt depth4.5pt width0pt}
\global\newcount\ftno \global\ftno=0
\def\foot{\global\advance\ftno by1\def\foot@rg{\hyperref{}{footnote}%
{\the\ftno}{\the\ftno}\xdef\foot@rg{\noexpand\hyperdef\noexpand\hypernoname%
{footnote}{\the\ftno}{\the\ftno}}}\footnote{$^{\foot@rg}$}}
%
%
%
\global\newcount\refno \global\refno=1
\newwrite\rfile
\def\ref{[\hyperref{}{reference}{\the\refno}{\the\refno}]\nref}
\def\nref#1{\DefWarn#1%
\xdef#1{[\noexpand\hyperref{}{reference}{\the\refno}{\the\refno}]}%
\writedef{#1\leftbracket#1}%
\ifnum\refno=1\immediate\openout\rfile=\jobname.refs\fi
\chardef\wfile=\rfile\immediate\write\rfile{\noexpand\item{[\noexpand\hyperdef%
\noexpand\hypernoname{reference}{\the\refno}{\the\refno}]\ }%
\reflabeL{#1\hskip.31in}\pctsign}\global\advance\refno by1\findarg}
\def\findarg#1#{\begingroup\obeylines\newlinechar=`\^^M\pass@rg}
{\obeylines\gdef\pass@rg#1{\writ@line\relax #1^^M\hbox{}^^M}%
\gdef\writ@line#1^^M{\expandafter\toks0\expandafter{\striprel@x #1}%
\edef\next{\the\toks0}\ifx\next\em@rk\let\next=\endgroup\else\ifx\next\empty%
\else\immediate\write\wfile{\the\toks0}\fi\let\next=\writ@line\fi\next\relax}}
\def\striprel@x#1{} \def\em@rk{\hbox{}}
\def\lref{\begingroup\obeylines\lr@f}
\def\lr@f#1#2{\DefWarn#1\gdef#1{\let#1=\UNd@FiNeD\ref#1{#2}}\endgroup\unskip}
\def\semi{;\hfil\break}
\def\addref#1{\immediate\write\rfile{\noexpand\item{}#1}} 
\def\listrefs{\vfill\supereject\immediate\closeout\rfile\writestoppt
\baselineskip=\footskip\centerline{{\bf References}}\bigskip{\parindent=20pt%
\frenchspacing\escapechar=` \input \jobname.refs\vfill\eject}\nonfrenchspacing}
\def\startrefs#1{\immediate\openout\rfile=\jobname.refs\refno=#1}
\def\xref{\expandafter\xr@f}\def\xr@f[#1]{#1}
\def\refs#1{\count255=1[\r@fs #1{\hbox{}}]}
\def\r@fs#1{\ifx\UNd@FiNeD#1\message{reflabel \string#1 is undefined.}%
\nref#1{need to supply reference \string#1.}\fi%
\vphantom{\hphantom{#1}}{\let\hyperref=\relax\xdef\next{#1}}%
\ifx\next\em@rk\def\next{}%
\else\ifx\next#1\ifodd\count255\relax\xref#1\count255=0\fi%
\else#1\count255=1\fi\let\next=\r@fs\fi\next}
%

%
\newwrite\ffile\global\newcount\figno \global\figno=1
\def\fig{fig.~\hyperref{}{figure}{\the\figno}{\the\figno}\nfig}
\def\nfig#1{\DefWarn#1%
\xdef#1{fig.~\noexpand\hyperref{}{figure}{\the\figno}{\the\figno}}%
\writedef{#1\leftbracket fig.\noexpand~\xfig#1}%
\ifnum\figno=1\immediate\openout\ffile=\jobname.figs\fi\chardef\wfile=\ffile%
{\let\hyperref=\relax
\immediate\write\ffile{\noexpand\medskip\noexpand\item{Fig.\ %
\noexpand\hyperdef\noexpand\hypernoname{figure}{\the\figno}{\the\figno}. }
\reflabeL{#1\hskip.55in}\pctsign}}\global\advance\figno by1\findarg}
\def\xfig{\expandafter\xf@g}\def\xf@g fig.\penalty\@M\ {}
\def\figs#1{figs.~\f@gs #1{\hbox{}}}
\def\f@gs#1{{\let\hyperref=\relax\xdef\next{#1}}\ifx\next\em@rk\def\next{}\else
\ifx\next#1\xfig #1\else#1\fi\let\next=\f@gs\fi\next}
%
\def\figin{\epsfcheck\figin}\def\figins{\epsfcheck\figins}
\def\epsfcheck{\ifx\epsfbox\UnDeFiNeD
\message{(NO epsf.tex, FIGURES WILL BE IGNORED)}
\gdef\figin##1{\vskip2in}\gdef\figins##1{\hskip.5in}
\else\message{(FIGURES WILL BE INCLUDED)}%
\gdef\figin##1{##1}\gdef\figins##1{##1}\fi}
\def\DefWarn#1{}
\def\figinsert{\goodbreak\topinsert}
\def\ifig#1#2#3{\DefWarn#1\xdef#1{fig.~\the\figno}
\writedef{#1\leftbracket fig.\noexpand~\the\figno}%
\figinsert\figin{\centerline{#3}}
\smallskip
\leftskip=0pt \rightskip=0pt
\baselineskip12pt\noindent
{{\bf Fig.~\the\figno}\ \ninepoint #2}
\medskip
\global\advance\figno by1\par\endinsert}
\newwrite\lfile
{\escapechar-1\xdef\pctsign{\string\%}\xdef\leftbracket{\string\{}
\xdef\rightbracket{\string\}}\xdef\numbersign{\string\#}}
\def\writedefs{\immediate\openout\lfile=label.defs \def\writedef##1{%
{\let\hyperref=\relax\let\hyperdef=\relax\let\hypernoname=\relax
 \immediate\write\lfile{\string\def\string##1\rightbracket}}}}%
\def\writestop{\def\writestoppt{\immediate\write\lfile{\string\pageno
 \the\pageno\string\startrefs\leftbracket\the\refno\rightbracket
 \string\def\string\secsym\leftbracket\secsym\rightbracket
 \string\secno\the\secno\string\meqno\the\meqno}\immediate\closeout\lfile}}
\def\writestoppt{}\def\writedef#1{}

\def\seclab#1{\DefWarn#1%
\xdef #1{\noexpand\hyperref{}{section}{\the\secno}{\the\secno}}%
\writedef{#1\leftbracket#1}\wrlabeL{#1=#1}\par%
\nobreak\medskip\nobreak\noindent\ignorespaces}
\def\subseclab#1\par{\DefWarn#1%
\xdef #1{\noexpand\hyperref{}{subsection}{\the\secno.\the\subsecno}%
{\the\secno.\the\subsecno}}\writedef{#1\leftbracket#1}\wrlabeL{#1=#1}\par%
\nobreak\medskip\nobreak\noindent\ignorespaces}
\def\applab#1{\DefWarn#1%
\xdef #1{\noexpand\hyperref{}{appendix}{\secn@m}{\secn@m}}%
\writedef{#1\leftbracket#1}\wrlabeL{#1=#1}}
\newwrite\tfile \def\writetoca#1{}
\def\leaderfill{\leaders\hbox to 1em{\hss.\hss}\hfill}
\def\writetoc{\immediate\openout\tfile=\jobname.toc
   \def\writetoca##1{{\edef\next{\write\tfile{\noindent ##1
   \string\leaderfill{
   \string\hyperref{}{page}{\noexpand\number\pageno}%
   {\noexpand\number\pageno}} \par}}\next}}
}
\newread\ch@ckfile
\def\listtoc{\immediate\closeout\tfile\immediate\openin\ch@ckfile=\jobname.toc
\ifeof\ch@ckfile\message{no file \jobname.toc, no table of contents this pass}%
\else\closein\ch@ckfile\centerline{\bf Contents}\nobreak\medskip%
{\baselineskip=16pt\footnotefont\parskip=0pt\catcode`\@=11\input\jobname.toc
\catcode`\@=12\bigbreak\bigskip}\fi}
\catcode`\@=12 
\def\tenpoint{\def\rm{\fam0\tenrm}
\textfont0=\tenrm \scriptfont0=\sevenrm \scriptscriptfont0=\fiverm
\textfont1=\teni  \scriptfont1=\seveni  \scriptscriptfont1=\fivei
\textfont2=\tensy \scriptfont2=\sevensy \scriptscriptfont2=\fivesy
\textfont\itfam=\tenit \def\it{\fam\itfam\tenit}\def\footnotefont{\ninepoint}%
\textfont\bffam=\tenbf \def\bf{\fam\bffam\tenbf}\def\sl{\fam\slfam\tensl}\rm}
\font\ninerm=cmr9 \font\sixrm=cmr6 \font\ninei=cmmi9 \font\sixi=cmmi6
\font\ninesy=cmsy9 \font\sixsy=cmsy6 \font\ninebf=cmbx9
\font\nineit=cmti9 \font\ninesl=cmsl9 \skewchar\ninei='177
\skewchar\sixi='177 \skewchar\ninesy='60 \skewchar\sixsy='60
\def\ninepoint{\def\rm{\fam0\ninerm}
\textfont0=\ninerm \scriptfont0=\sixrm \scriptscriptfont0=\fiverm
\textfont1=\ninei \scriptfont1=\sixi \scriptscriptfont1=\fivei
\textfont2=\ninesy \scriptfont2=\sixsy \scriptscriptfont2=\fivesy
\textfont\itfam=\ninei \def\it{\fam\itfam\nineit}\def\sl{\fam\slfam\ninesl}%
\textfont\bffam=\ninebf \def\bf{\fam\bffam\ninebf}\rm}
%
\hyphenation{anom-aly anom-alies coun-ter-term coun-ter-terms}

\global\newcount\subsubsecno \global\subsubsecno=0
\def\subsubsec#1\par{\global\advance\subsubsecno by1%
{\toks0{#1}\message{(\the\secno\the\subsecno\the\subsubsecno. \the\toks0)}}%
\ifnum\lastpenalty>9000\else\bigbreak\fi
\noindent{\it\hyperdef\hypernoname{subsubsection}{\the\secno.\the\subsecno\the\subsubsecno}%
{\the\secno.\the\subsecno.\the\subsubsecno.} #1}
\par\nobreak\medskip\nobreak\noindent\ignorespaces}

\def\DefWarn#1{}
\def\tikzcaption#1#2{\DefWarn#1\xdef#1{Fig.~\the\figno}
\writedef{#1\leftbracket Fig.\noexpand~\the\figno}%
{
\smallskip
\leftskip=20pt \rightskip=20pt \baselineskip12pt\noindent
{{\bf Fig.~\the\figno}\ \ninepoint #2}
\bigskip
\global\advance\figno by1 \par}}

\def\ntoalpha#1{%
\ifcase#1%
@%
\or A\or B\or C\or D\or E\or F\or G\or H\or I
\fi
}

\global\newcount\appno \global\appno=1
\def\applab#1{\xdef #1{\ntoalpha\appno}\writedef{#1\leftbracket#1}\wrlabeL{#1=#1}
\global\advance\appno by1}

\def\preprint#1 #2\par{\rightline{\vbox{\baselineskip12pt\hbox{#1}\hbox{#2}}}\vskip2cm}
%
\def\title#1\par{\centerline{\bf #1}\nopagenumbers\pageno=0}
\def\author#1\par{\bigskip\bigskip\centerline{#1}}

\newcount\addressno

\def\email#1#2{\unskip$^#1$\footnote{\null}{\kern-\parindent \llap{$^#1$\hskip1pt}email: #2}}

\def\startcenter{%
  \par
  \begingroup
  \leftskip=0pt plus 1fil
  \rightskip=\leftskip
  \parindent=0pt
  \parfillskip=0pt
}
\def\stopcenter{\endgroup}

\def\address{\bigskip%
  \ifnum\the\addressno=0\else\stopcenter\endgroup\fi
  \advance\addressno by 1%
  \begingroup
  \startcenter
  \it
  \obeylines
  \addressAux
}
\def\addressAux#1{#1}

\def\abstract{\stopcenter\endgroup\bigskip\bigskip\noindent}

\def\Dsl{\,\raise.15ex\hbox{/}\mkern-13.5mu D} 
\def\dsl{\raise.15ex\hbox{/}\kern-.57em\partial}
\def\tr{{\rm tr}} 
\def\boxeqn#1{\vcenter{\vbox{\hrule\hbox{\vrule\kern3pt\vbox{\kern3pt
	\hbox{${\displaystyle #1}$}\kern3pt}\kern3pt\vrule}\hrule}}}


\def\ap{{\alpha^{\prime}}}

\def\a{\alpha}

\def\g{{\gamma}}
\def\d{{\delta}}

\def\l{\lambda}

\def\t{{\theta}}

\def\half{{1\over 2}}
\def\p{{\partial}}

\def\bar{\overline}
\def\({\left(}
\def\){\right)}
\def\cF{{\cal F}}
\def\cW{{\cal W}}
\def\cY{{\cal Y}}
\def\cA{{\cal A}}

\def\Im{\mathop{{\rm Im}}} 
\def\sfrac#1/#2{\kern.1em\raise.5ex\hbox{\the\scriptfont0 #1}%
\kern-.1em/\kern-.15em\lower.25ex\hbox{\the\scriptfont0 #2}}



\def\qed{\hbox{\hskip 3pt
\vbox{\hrule\hbox to 7pt{\vrule height 7pt\hfill\vrule}
\hrule}}\hskip3pt}

\overfullrule=0pt\relax

\frenchspacing

\newread\instream \openin\instream= label.defs
\ifeof\instream \message{No labels in advance yet. Wait till next pass.}
\else \closein\instream \input label.defs
\fi
\writedefs

\def\arXiv:#1].{\hepthStrip#1 \nil}
\def\hepthStrip#1 #2\nil{\href{http://arxiv.org/abs/#1}{arXiv:#1 #2\unskip}].}

\def\cK{{\cal K}}
\def\dd{{\rm d}}
\def\vep{\varepsilon}
\def\frac#1#2{{#1\over #2}}
\def\te#1{{\rm #1}}

\phantom{A}
\vskip2cm

\title One-loop superstring six-point amplitudes

\title and anomalies in pure spinor superspace

\author
Carlos R. Mafra\email{\star}{mafra@ias.edu} and
Oliver Schlotterer\email{\dagger}{olivers@aei.mpg.de}

\address
$^\star$Institute for Advanced Study, School of Natural Sciences,
Einstein Drive, Princeton, NJ 08540, USA
\medskip
$^\dagger$Max--Planck--Institut f\"ur Gravitationsphysik,
Albert--Einstein--Institut,
Am Muehlenberg, 14476 Potsdam, Germany

\abstract
We present the massless six-point one-loop amplitudes in the open and closed
superstring using BRST cohomology arguments from the pure spinor formalism. 
The hexagon gauge anomaly is traced back to a class of kinematic factors in pure
spinor superspace which were recently introduced as BRST pseudo-invariants.
This complements previous work where BRST invariance arguments were used to derive
the non-anomalous part of the amplitude.
The associated worldsheet functions are non-singular and demonstrated to yield
total derivatives on moduli space upon gauge variation. These cohomology considerations
yield an efficient organizing principle for closed-string amplitudes that match expectations
from S-duality in the low-energy limit.

\Date {March 2016}


\lref\Mafrakh{
  C.R.~Mafra and O.~Schlotterer,
  ``The Structure of n-Point One-Loop Open Superstring Amplitudes,''
  JHEP {\bf 1408} (2014) 099
  [arXiv:1203.6215 [hep-th]].

}\lref\PSanomaly{
  N.~Berkovits and C.R.~Mafra,
  ``Some Superstring Amplitude Computations with the Non-Minimal Pure Spinor Formalism,''
  JHEP {\bf 0611}, 079 (2006)
  [hep-th/0607187].

}\lref\NMPS{
  N.~Berkovits,
  ``Pure spinor formalism as an N=2 topological string,''
  JHEP {\bf 0510} (2005) 089
  [hep-th/0509120].

}\lref\Greenqs{
  M.~B.~Green and J.~H.~Schwarz,
  ``The Hexagon Gauge Anomaly in Type I Superstring Theory,''
  Nucl.\ Phys.\ B {\bf 255} (1985) 93.
\semi
  M.~B.~Green and J.~H.~Schwarz,
  ``Anomaly Cancellation in Supersymmetric D=10 Gauge Theory and Superstring Theory,''
  Phys.\ Lett.\ B {\bf 149} (1984) 117.
}
\lref\FMS{
  D.~Friedan, E.~J.~Martinec and S.~H.~Shenker,
  ``Conformal Invariance, Supersymmetry and String Theory,''
  Nucl.\ Phys.\ B {\bf 271} (1986) 93.
}
\lref\fiveptNMPS{
  C.R.~Mafra and C.~Stahn,
  ``The One-loop Open Superstring Massless Five-point Amplitude with the Non-Minimal Pure Spinor Formalism,''
  JHEP {\bf 0903} (2009) 126
  [arXiv:0902.1539 [hep-th]].
}
\lref\Berkovitsfe{
  N.~Berkovits,
  ``Super Poincare covariant quantization of the superstring,''
  JHEP {\bf 0004} (2000) 018
  [hep-th/0001035].
}
\lref\Richardsjg{
  D.~M.~Richards,
  ``The One-Loop Five-Graviton Amplitude and the Effective Action,''
  JHEP {\bf 0810} (2008) 042
  [arXiv:0807.2421 [hep-th]].
 
}\lref\EOMBBs{
  C.R.~Mafra and O.~Schlotterer,
  ``Multiparticle SYM equations of motion and pure spinor BRST blocks,''
  JHEP {\bf 1407} (2014) 153
  [arXiv:1404.4986 [hep-th]].

}
\lref\cohom{
  C.R.~Mafra and O.~Schlotterer,
  ``Cohomology foundations of one-loop amplitudes in pure spinor superspace,''
  arXiv:1408.3605 [hep-th].
}
\lref\emzv{
J.~Broedel, C.R.~Mafra, N.~Matthes and O.~Schlotterer,
  ``Elliptic multiple zeta values and one-loop superstring amplitudes,''
JHEP {\bf 1507}, 112 (2015).
[arXiv:1412.5535 [hep-th]].

}\lref\Mafrajq{
  C.R.~Mafra, O.~Schlotterer, S.~Stieberger and D.~Tsimpis,
  ``A recursive method for SYM n-point tree amplitudes,''
  Phys.\ Rev.\ D {\bf 83} (2011) 126012
  [arXiv:1012.3981 [hep-th]].
}

\lref\Mafranv{
  C.R.~Mafra, O.~Schlotterer and S.~Stieberger,
  ``Complete N-Point Superstring Disk Amplitude I. Pure Spinor Computation,''
  Nucl.\ Phys.\ B {\bf 873} (2013) 419
  [arXiv:1106.2645 [hep-th]].
}
\lref\Barreirodpa{
  L.~A.~Barreiro and R.~Medina,
``RNS derivation of N-point disk amplitudes from the revisited S-matrix approach,''
Nucl.\ Phys.\ B {\bf 886}, 870 (2014).
[arXiv:1310.5942 [hep-th]].

}\lref\Barreiroaw{
  L.~A.~Barreiro and R.~Medina,
  ``Revisiting the S-matrix approach to the open superstring low energy effective lagrangian,''
  JHEP {\bf 1210} (2012) 108
  [arXiv:1208.6066 [hep-th]].
}

\lref\MPS{
  N.~Berkovits,
  ``Multiloop amplitudes and vanishing theorems using the pure spinor formalism for the superstring,''
  JHEP {\bf 0409}, 047 (2004)
  [hep-th/0406055].

}
\lref\twoloop{
  N.~Berkovits,
  ``Super-Poincare covariant two-loop superstring amplitudes,''
JHEP {\bf 0601}, 005 (2006).
[hep-th/0503197].
\semi
  N.~Berkovits and C.R.~Mafra,
  ``Equivalence of two-loop superstring amplitudes in the pure spinor and RNS formalisms,''
Phys.\ Rev.\ Lett.\  {\bf 96}, 011602 (2006).
[hep-th/0509234].
}
\lref\fourptoneloop{
  C.R.~Mafra,
  ``Four-point one-loop amplitude computation in the pure spinor formalism,''
JHEP {\bf 0601}, 075 (2006).
[hep-th/0512052].
}

\lref\OdaSD{
  I.~Oda and M.~Tonin,
  ``Y-formalism in pure spinor quantization of superstrings,''
Nucl.\ Phys.\ B {\bf 727}, 176 (2005).
[hep-th/0505277].
}

\lref\Tsuchiyava{
  A.~Tsuchiya,
  ``More on One Loop Massless Amplitudes of Superstring Theories,''
  Phys.\ Rev.\ D {\bf 39} (1989) 1626.

}\lref\Stiebergerwk{
  S.~Stieberger and T.~R.~Taylor,
  ``NonAbelian Born-Infeld action and type I. Heterotic duality (1): Heterotic F**6 terms at two loops,''
  Nucl.\ Phys.\ B {\bf 647}, 49 (2002)
  [hep-th/0207026].
  S.~Stieberger and T.~R.~Taylor,
  ``NonAbelian Born-Infeld action and type 1. - heterotic duality 2: Nonrenormalization theorems,''
  Nucl.\ Phys.\ B {\bf 648} (2003) 3
  [hep-th/0209064].

}\lref\Clavellifj{
  L.~Clavelli, P.~H.~Cox and B.~Harms,
  ``Parity Violating One Loop Six Point Function in Type I Superstring Theory,''
  Phys.\ Rev.\ D {\bf 35} (1987) 1908.

}\lref\Greenbza{
  M.~B.~Green, C.R.~Mafra and O.~Schlotterer,
  ``Multiparticle one-loop amplitudes and S-duality in closed superstring theory,''
  JHEP {\bf 1310} (2013) 188
  [arXiv:1307.3534].

}\lref\BjerrumBohrvc{ 
  N.~E.~J.~Bjerrum-Bohr and P.~Vanhove,
  ``Explicit Cancellation of Triangles in One-loop Gravity Amplitudes,''
  JHEP {\bf 0804}, 065 (2008)
  [arXiv:0802.0868 [hep-th]].

}\lref\Greenpv{
  M.~B.~Green and P.~Vanhove,
  ``The Low-energy expansion of the one loop type II superstring amplitude,''
  Phys.\ Rev.\ D {\bf 61} (2000) 104011
  [hep-th/9910056].

}\lref\Greenuj{
  M.~B.~Green, J.~G.~Russo and P.~Vanhove,
  ``Low energy expansion of the four-particle genus-one amplitude in type II superstring theory,''
  JHEP {\bf 0802} (2008) 020
  [arXiv:0801.0322 [hep-th]].
  
}\lref\Greensw{
  M.~B.~Green, J.~H.~Schwarz and L.~Brink,
  ``N=4 Yang-Mills and N=8 Supergravity as Limits of String Theories,''
  Nucl.\ Phys.\ B {\bf 198} (1982) 474.

}\lref\Greentv{
  M.~B.~Green and M.~Gutperle,
  ``Effects of D instantons,''
  Nucl.\ Phys.\ B {\bf 498} (1997) 195
  [hep-th/9701093].
    
}\lref\Greenby{
  M.~B.~Green and S.~Sethi,
  ``Supersymmetry constraints on type IIB supergravity,''
  Phys.\ Rev.\ D {\bf 59} (1999) 046006
  [hep-th/9808061].

}\lref\Sinhazr{
  A.~Sinha,
  ``The $\hat G^4 \lambda^{16}$ term in IIB supergravity,''
  JHEP {\bf 0208} (2002) 017
  [hep-th/0207070].
  
}\lref\Schlottererny{
  O.~Schlotterer and S.~Stieberger,
``Motivic Multiple Zeta Values and Superstring Amplitudes,''
J.\ Phys.\ A {\bf 46}, 475401 (2013).
[arXiv:1205.1516 [hep-th]].
  
}\lref\BjerrumBohrhn{
  N.~E.~J.~Bjerrum-Bohr, P.~H.~Damgaard, T.~Sondergaard and P.~Vanhove,
  ``The Momentum Kernel of Gauge and Gravity Theories,''
  JHEP {\bf 1101} (2011) 001
  [arXiv:1010.3933 [hep-th]].
  
}\lref\Kawaixq{
  H.~Kawai, D.~C.~Lewellen and S.~H.~H.~Tye,
  ``A Relation Between Tree Amplitudes of Closed and Open Strings,''
  Nucl.\ Phys.\ B {\bf 269} (1986) 1.
  
}\lref\Bernsv{
  Z.~Bern, L.~J.~Dixon, M.~Perelstein and J.~S.~Rozowsky,
  ``Multileg one loop gravity amplitudes from gauge theory,''
  Nucl.\ Phys.\ B {\bf 546} (1999) 423
  [hep-th/9811140].

}\lref\WWW{ J. Broedel, O.~Schlotterer and S.~Stieberger,
{\tt http://mzv.mpp.mpg.de}
}

\lref\AWeyl{
A. Weil, ``Elliptic Functions according to Eisenstein and Kronecker'', Springer-Verlag, 1976.
}

\lref\Broedelaza{
  J.~Broedel, O.~Schlotterer, S.~Stieberger and T.~Terasoma,
  ``All order $\alpha^{\prime}$-expansion of superstring trees from the Drinfeld associator,''
Phys.\ Rev.\ D {\bf 89}, no. 6, 066014 (2014).
[arXiv:1304.7304 [hep-th]].
}

\lref\Oprisawu{
  D.~Oprisa and S.~Stieberger,
  ``Six gluon open superstring disk amplitude, multiple hypergeometric series and Euler-Zagier sums,''
  hep-th/0509042.

}\lref\Stiebergerbh{
  S.~Stieberger and T.~R.~Taylor,
  ``Amplitude for N-Gluon Superstring Scattering,''
  Phys.\ Rev.\ Lett.\  {\bf 97} (2006) 211601
  [hep-th/0607184].
  
}\lref\Stiebergerte{
  S.~Stieberger and T.~R.~Taylor,
  ``Multi-Gluon Scattering in Open Superstring Theory,''
  Phys.\ Rev.\ D {\bf 74} (2006) 126007
  [hep-th/0609175].
}

\lref\Drummondvz{
  J.~M.~Drummond and E.~Ragoucy,
``Superstring amplitudes and the associator,''
JHEP {\bf 1308}, 135 (2013).
[arXiv:1301.0794 [hep-th]].
  
}
\lref\Boelsjua{
  R.~H.~Boels,
``On the field theory expansion of superstring five point amplitudes,''
Nucl.\ Phys.\ B {\bf 876}, 215 (2013).
[arXiv:1304.7918 [hep-th]].
}
\lref\Boelszr{
  R.~H.~Boels,
  ``Maximal R-symmetry violating amplitudes in type IIB superstring theory,''
  Phys.\ Rev.\ Lett.\  {\bf 109} (2012) 081602
  [arXiv:1204.4208 [hep-th]].
}
\lref\wittentwistor{ 
  E.~Witten,
  ``Twistor - Like Transform in Ten-Dimensions,''
  Nucl.\ Phys.\ B {\bf 266}, 245 (1986).
}

\lref\symanom{
  C.R.~Mafra and O.~Schlotterer,
  ``Towards one-loop SYM amplitudes from the pure spinor BRST cohomology,''
Fortsch.\ Phys.\  {\bf 63}, no. 2, 105 (2015).
[arXiv:1410.0668 [hep-th]].
}

\lref\psweb{
  C.R.~Mafra, O.~Schlotterer,
http://www.damtp.cam.ac.uk/user/crm66/SYM/pss.html

}\lref\Grosspd{
  D.~J.~Gross and P.~F.~Mende,
  ``Modular Subgroups, Odd Spin Structures and Gauge Invariance in the Heterotic String,''
  Nucl.\ Phys.\ B {\bf 291} (1987) 653.
}

\lref\GomezSLA{
  H.~Gomez and C.R.~Mafra,
  ``The closed-string 3-loop amplitude and S-duality,''
JHEP {\bf 1310}, 217 (2013).
[arXiv:1308.6567 [hep-th]].
}

\lref\GomezUHA{
  H.~Gomez, C.R.~Mafra and O.~Schlotterer,
``Two-loop superstring five-point amplitude and S-duality,''
Phys.\ Rev.\ D {\bf 93}, no. 4, 045030 (2016).
[arXiv:1504.02759 [hep-th]].
}

\lref\MafraMJA{
  C.R.~Mafra and O.~Schlotterer,
  ``Two-loop five-point amplitudes of super Yang-Mills and supergravity in pure spinor superspace,''
JHEP {\bf 1510}, 124 (2015).
[arXiv:1505.02746 [hep-th]].
}

\lref\MafraVCA{
 C.~R.~Mafra and O.~Schlotterer,
  ``Berends-Giele recursions and the BCJ duality in superspace and components,''
JHEP {\bf 1603}, 097 (2016).
[arXiv:1510.08846 [hep-th]].
}

\lref\LeeUPY{
 S.~Lee, C.~R.~Mafra and O.~Schlotterer,
  ``Non-linear gauge transformations in $D=10$ SYM theory and the BCJ duality,''
JHEP {\bf 1603}, 090 (2016).
[arXiv:1510.08843 [hep-th]].
}

\lref\MafraGIA{
  C.R.~Mafra and O.~Schlotterer,
  ``Solution to the nonlinear field equations of ten dimensional supersymmetric Yang-Mills theory,''
Phys.\ Rev.\ D {\bf 92}, no. 6, 066001 (2015).
[arXiv:1501.05562 [hep-th]].
}

\lref\BBS{
M.~Berg, I.~Buchberger and O.~Schlotterer,
  ``From maximal to minimal supersymmetry in string loop amplitudes,''
[arXiv:1603.05262 [hep-th]].
}

\lref\Cai{
  J.~Polchinski and Y.~Cai,
  ``Consistency of Open Superstring Theories,''
Nucl.\ Phys.\ B {\bf 296}, 91 (1988).
}
\lref\Hayashi{
  M.~Hayashi, N.~Kawamoto, T.~Kuramoto and K.~Shigemoto,
  ``Modular Invariance and Gravitational Anomaly in Type II Superstring Theory,''
Nucl.\ Phys.\ B {\bf 294}, 459 (1987).
}
\lref\Kutasov{
  D.~Kutasov,
  ``Modular Invariance, Chiral Anomalies and Contact Terms,''
Nucl.\ Phys.\ B {\bf 307}, 417 (1988).
}
\lref\LercheNP{
  W.~Lerche, A.~N.~Schellekens and N.P.~Warner,
  ``Lattices and Strings,''
Phys.\ Rept.\  {\bf 177}, 1 (1989).
}
\lref\Kubota{
  T.~Inami, H.~Kanno and T.~Kubota,
  ``Hexagon Gauge Anomaly and Supermoduli in the Path Integral Method of Superstrings,''
Nucl.\ Phys.\ B {\bf 308}, 203 (1988).
}
\lref\BerkovitsRPA{
  N.~Berkovits and E.~Witten,
  ``Supersymmetry Breaking Effects using the Pure Spinor Formalism of the Superstring,''
JHEP {\bf 1406}, 127 (2014).
[arXiv:1404.5346 [hep-th]].
}
\lref\HeWGF{
  S.~He, R.~Monteiro and O.~Schlotterer,
  ``String-inspired BCJ numerators for one-loop MHV amplitudes,''
JHEP {\bf 1601}, 171 (2016).
[arXiv:1507.06288 [hep-th]].
}

\listtoc
\writetoc
\filbreak

\newsec Introduction

Over the past decade, several superstring \refs{\MPS,\twoloop, \Mafranv, \GomezSLA, \GomezUHA} 
and field-theory scattering amplitudes \refs{\Mafrajq, \symanom, \MafraMJA} have been
computed in manifestly supersymmetric form using the pure spinor formalism \Berkovitsfe.
Computations in the minimal pure spinor formalism relied extensively on the BRST invariance of the amplitude prescription
as a way to organize the intermediate steps and to simplify the answers. At tree
level, this method led to a general solution in closed form for the $n$-point
integrand for both the open superstring \Mafranv\ as well as its field-theory limit \Mafrajq. At
higher loops --- apart from the four-point one- and two-loop amplitudes 
of~\refs{\MPS,\fourptoneloop,\twoloop} --- the superstring computations of 
\refs{\GomezSLA, \GomezUHA} so far were restricted to the low-energy limit
of the integrand. This limit only receives contributions
from a subset of the zero-modes of the pure spinor b-ghost and leads to a simpler analysis
of OPE singularities among external vertex operators.

In 2012 \Mafrakh, the one-loop open superstring $n$-point integrand restricted to
the above zero-mode contributions of the b-ghost was computed in closed form in
terms of scalar BRST invariants denoted by $C_{i|A,B,C}$. These BRST invariants
were later given a recursive construction in terms of ten-dimensional SYM
superfields including a general expansion in terms of field-theory tree amplitudes \EOMBBs.
Although the permutation-invariant integrands in \Mafrakh\ yield the desired
low-energy behavior, they fail to reproduce the hexagon gauge anomaly on the
boundary of moduli space.

The long-term goal of this project is to lift the restriction of b-ghost zero-modes from
the one-loop analysis of \Mafrakh\ in order to obtain the complete and
supersymmetric $n$-point one-loop amplitudes of the open superstring. In this
paper we take the first step and write the complete six-point
one-loop integrand for open and closed superstrings in pure spinor superspace.
These results reproduce the pure spinor analysis of the gauge
anomaly in \PSanomaly\ and match previous computations done with the RNS formalism.
But unlike the RNS answer which is restricted
to gluon amplitudes (see \Tsuchiyava\ for the parity-even and
\refs{\Clavellifj,\Kubota} for the parity-odd part), the result of this paper is
fully supersymmetric and naturally unifies the contributions from both the even and
the odd spin structures. Moreover, the worldsheet integrals for both
open and closed strings are cast into a basis.
For closed strings, a new 
factorized representation of the five-point kinematics paves the way for an
efficient organization of the six-point result.

Since the gauge anomaly
probes non-standard contributions from the b-ghost beyond the zero-mode analysis of
\Mafrakh, the six-point one-loop result of this paper harbors
important insights about a difficult corner of the pure spinor formalism which
currently inhibits further progress in multiloop computations.

\newsec{Review: the hexagon anomaly and its cancellation}

\subsec The pure spinor description of the anomalous gauge variation
\par\subseclab\sectwotwo

\noindent The gauge variation of the six-point open-superstring amplitude at one loop
using the pure spinor formalism was computed in \PSanomaly. This subsection briefly
reviews that derivation.

The non-minimal pure spinor prescription to compute a one-loop amplitude
in the type-I superstring with
a $SO(N)$ gauge group is given by \NMPS
\eqn\nmpsAmp{
{\cal A}_n = \sum_{\rm top} G_{{\rm top}}\int_0^{\infty} \dd t
\int_{\Delta_{\rm top}}\!\!\!\!\!  \dd z_2 \, \dd z_3 \, \ldots \, \dd z_n \,\, \langle{ {\cal N} (b,\mu)
    V_1   \prod_{j=2}^n U_j (z_j)}\rangle \ .
}
The sum is over the three worldsheet topologies at one-loop with $G_{\rm top}$
and $\Delta_{\rm top}$ denoting their corresponding Chan--Paton factors and
integration domains for $z_j$. Denoting the generators of $SO(N)$ in the
fundamental representation by $t^{a_i}$, the Chan--Paton factors for the cylinder with
all particles attached to one boundary and the M\"obius
strip are given by
$G_P=N \tr{(t^{a_1}t^{a_2}t^{a_3}t^{a_4}t^{a_5}t^{a_6})}$
and $G_{N}=-{\rm tr}{(t^{a_1}t^{a_2}t^{a_3}t^{a_4}t^{a_5}t^{a_6})}$.
When particles are attached to
both boundaries of the cylinder one has, for example, $G_{NP}={\rm
tr}{(t^{a_1}t^{a_2})}{\rm tr}{(t^{a_3}t^{a_4}t^{a_5}t^{a_6})}$.
The integration domains will be
elaborated in section~\secfourone.

Furthermore, $t$ is the one-loop Teichm\"uller parameter and $\mu$ the Beltrami
differential, $b$ is the b-ghost (see \NMPS\ for the expression in the non-minimal formalism and \refs{\MPS, \OdaSD} for its schematic form in the minimal formalism), 
and $(b,\mu) \equiv \int \dd^2 w\, b(w)\mu$. The massless vertices are \Berkovitsfe\
\eqn\vertices{
V = \lambda^\a A_\a,\qquad U= \p\theta^\a A_\a + \Pi^m A_m + d_\a W^\a + {1\over 2}N^{mn}F_{mn}
}
with pure spinor $\lambda^\alpha$ subject to $(\lambda \gamma^m \lambda)=0$, 
linearized superfields $[A_\a ,A_m , W^\a , F_{mn}]$ of ten-dimensional SYM
\wittentwistor\ and worldsheet fields $[ \p\theta^\a,\Pi^m, d_\a ,N^{mn}]$ of
conformal weight $h=1$ whose OPEs can be found in \Berkovitsfe. Finally, ${\cal N}$
regulates the integration over the non-compact space of pure spinors \NMPS.

As in the original derivation of \Greenqs, the gauge variation of the amplitude
can be computed directly by replacing the vertex operators by their gauge
variation
\eqn\gaugeVar{
\delta V_1 = Q\Omega_1 \ , \ \ \ \ \ \ 
\delta U_2 = \partial \Omega_2
\,,
}
where $\Omega_j$ are scalar superfields, and the BRST charge is defined by
\eqn\Qch{
Q \equiv \lambda^\alpha D_\alpha \ , \ \ \ \ 
D_\alpha \equiv \frac{ \partial }{\partial \theta^\alpha} + {1\over 2} (\gamma_m \theta)_\alpha k^m
\, .
}
Since the total derivatives $\partial \Omega_2 \equiv {\partial \Omega_2 \over \partial z_2}$ from the integrated vertex operators are suppressed by the boundary contribution $z_i \rightarrow z_j$ of the integrand, the gauge variation of the six-point amplitude becomes
\eqnn\deltaAmp
$$\eqalignno{
\delta{\cal A}_6 &= \sum_{\rm top} G_{\rm top}\int_0^{\infty} \dd t \int_{\Delta_{\rm top}} \dd z_2 \, \ldots \, \dd z_6 \, \langle{ {\cal N} (b,\mu)
    (Q\Omega_1) \prod_{j=2}^6  U_j (z_j)}\rangle  &\deltaAmp\cr
&= -\sum_{\rm top} G_{\rm top}\int_0^{\infty} \dd t  {\partial \over \partial t} \int_{\Delta_{\rm top}} \dd z_2 \, \ldots \, \dd z_6 \,  \langle{ {\cal N}
    \Omega_1  \prod_{j=2}^6 U_j (z_j)}\rangle\,.
}$$
To arrive at the second line the BRST charge was integrated by parts. The only
non-vanishing contribution comes from the energy momentum tensor $\{Q,b\} = T$ and gives rise to a factor of
$(T,\mu)$ which in turn leads to a total derivative ${\partial\over\partial t}$
on moduli space \FMS.

The correlator in the second line of \deltaAmp\ can be easily evaluated by considering the
saturation of fermionic zero-modes of the fermionic field $d_\alpha$. It is
well known \NMPS\ that at one loop the regulator ${\cal N}$ provides eleven
zero-modes of $d_\alpha$, so the vertices contribute the remaining five in order
for the variation \deltaAmp\ to be non-vanishing,
$(dW_2)(dW_3)(dW_4)(dW_5)(dW_6)$. Integrating the pure spinor zero-modes has the
effect of replacing
$d_{\alpha_1}d_{\alpha_2}d_{\alpha_3}d_{\alpha_4}d_{\alpha_5} \rightarrow (\lambda
\g^m)_{\alpha_1}(\lambda \g^n)_{\alpha_2}(\lambda \g^p)_{\alpha_3}(\g_{mnp})_{\alpha_4\alpha_5}$ \fiveptNMPS,
and \deltaAmp\ becomes
\eqn\deltaAmptwo{
\delta{\cal A}_6 \sim K \sum_{\rm top} G_{\rm top}
\int_{\Delta_{\rm top}}\!\!\! \dd z_2 \, \ldots \, \dd z_6 \,
\Big \langle \prod_{j=1}^{n} e^{i k_j \cdot x(z_j,\bar z_j)} \Big \rangle
\big|^{t\rightarrow \infty}_{t\rightarrow 0}
}
with the following kinematic factor for the hexagon gauge anomaly \PSanomaly
\eqn\gaugevar{
K \equiv \langle\Omega_1 (\lambda\g^m W_2)(\lambda\g^n W_3)(\lambda\g^p W_4) (W_5\g_{mnp}W_6)\rangle \ .
}
The standard correlator of plane waves $e^{i k_j \cdot x(z_j,\bar z_j)}$ is detailed in section \secthreeone. The
component expansion of \gaugevar\ can be computed using the zero-mode integration prescription \Berkovitsfe
\eqn\zerom{
\langle (\l \g^m
\theta) (\l \g^n \theta) (\l \g^p \theta)
(\theta \g_{mnp} \theta) \rangle =2880
}
and, when restricted to gluonic fields with polarization vectors $e_i$, is proportional
to $\epsilon_{m_1 n_1 \ldots m_5 n_5} k_2^{m_1} e_2^{n_1} \ldots  k_6^{m_5} e_6^{n_5}$. 
In the next sections the result \deltaAmptwo\ will be re-derived from the
gauge variation of an explicit expression for the six-point amplitude at one loop.

\subsec Multiparticle kinematic building blocks
\par\subseclab\secmultpar

The zero-mode structure of the six-point one-loop amplitude in the pure spinor 
formalism \nmpsAmp\ allows for two OPEs 
among massless vertex operators. Such OPEs
can be recursively addressed using non-local multiparticle superfields
${\cal K}_P \in \{ \cA^P_\alpha$, $\cA_P^m$, $\cW_P^\alpha,\cF^{mn}_P \}$ of ten dimensional SYM \EOMBBs.
They are referred to as Berends--Giele currents and defined by
\eqn\BGdef{
\cK_P \equiv {1\over s_{P}}\sum_{XY=P}\cK_{[X,Y]} \,, 
}
where the multiparticle label $P=12\ldots p$ encompasses $p$ external legs. The sum in \BGdef\
instructs to deconcatenate $P$ into non-empty words $X=12\ldots j$ and $Y=j+1\ldots p$ with $j=1,2,\ldots,p-1$. 
The shorthand $\cK_{[X,Y]}$ is used to represent all the four
types of superfields simultaneously. More explicitly \LeeUPY,
\eqnn\PPseven
\eqnn\PPeight
\eqnn\PPnine
\eqnn\PPten
$$\eqalignno{
\cA^{[P,Q]}_\a &\equiv - \half\bigl[  \cA^{P}_\a (k^{P}\cdot  \cA^Q)
+  \cA^{P}_m (\g^m \cW^Q)_\a - (P\leftrightarrow Q)\bigr]  &\PPseven\cr
\cA^{[P,Q]}_m &\equiv - \half\bigl[ \cA^{P}_m (k^P\cdot  \cA^{Q}) + \cA^{P}_n  \cF^Q_{mn}
- ( \cW^{P}\g_m \cW^Q)
- (P \leftrightarrow Q)\bigr] &\PPeight\cr
\cW_{[P,Q]}^\a &\equiv  \half (k_P^m + k_Q^m) \gamma_m^{\alpha \beta}
 \big[ \cA_P^n (\gamma_n \cW_Q)_\beta  - (P \leftrightarrow Q) \big]
&\PPnine\cr
\cF^{mn}_P &\equiv k_P^m \cA_P^n - k_P^n \cA_P^m
- \sum_{XY=P}\!\!\big( \cA_X^m \cA_Y^n - \cA_X^n \cA_Y^m \big) \ .
&\PPten\cr
}$$
Multiparticle momenta for $P=12\ldots p$ and their associated Mandelstam invariants are given by
\eqn\multmand{
k_P^m \equiv k^m_{1}+k^m_{2}+\cdots + k^m_{p}
\ , \ \ \ \ \ \
s_P\equiv{1\over 2}k_P^2\, .
}
Furthermore, we define the multiparticle version of the vertex operator $V$ in \vertices\ as
\eqn\McalVdef{
M_P \equiv \l^\a \cA^P_\a\,,
}
such that $M_i = V_i$. The zero-mode saturation in the pure spinor
one-loop amplitude prescription selects certain superfields
from the integrated vertex operators $U$ in \vertices, such as $V_1 (\l \g_m W_2)(\l \g_n W_3) F^{mn}_4$ 
in the four-point amplitude \MPS. Promoting the superfields to their 
Berends--Giele currents such as $W_i^\alpha \rightarrow {\cal W}_A^{\alpha}$ suggests the following definitions \refs{\EOMBBs, \cohom},
\eqnn\PPtwelve
\eqnn\PPthirteen
\eqnn\PPfourteen
\eqnn\PPfifteen
$$\eqalignno{
M_{A,B,C}&\equiv {1\over 3} (\l \g_m {\cal W}_A)(\l \g_n {\cal W}_B) {\cal F}^{mn}_C + (A\leftrightarrow B,C)
&\PPtwelve \cr
\cW^m_{A,B,C,D} &\equiv  {1\over 12}\bigl[ (\cW_A \g^{mnp} \cW_B)(\l \g _n \cW_C)
(\l \g_p \cW_D) + (A,B|A,B,C,D)\bigr] &\PPthirteen \cr
M^m_{A,B,C,D} &\equiv  \cW^m_{A,B,C,D} + \bigl[  \cA_A^m M_{B,C,D}
+ (A\leftrightarrow B,C,D)\bigr] &\PPfourteen \cr
M^{mn}_{A,B,C,D,E}&\equiv
\cA^m_A \cW^n_{B,C,D,E} + \cA^n_A M^m_{B,C,D,E} + (A \leftrightarrow B,C,D,E) \, ,
&\PPfifteen
}$$
which automatically capture the results of iterated OPEs. In \PPthirteen\ and later places, 
the notation $(a_1,{\ldots } , a_p \,|\, a_1,{\ldots} ,a_n)$
instructs to sum over all possible ways to choose $p$ elements $a_1,a_2,\ldots
,a_p$ out of the set $\{a_1,{\ldots} ,a_n\}$, for a total of ${n\choose p}$ terms.

\subsec BRST invariants
\par\subseclab\reviewBRST

The zero-mode bracket in \zerom\ which picks up the unique scalar of order
$\lambda^3 \theta^5$ from the enclosed superfields converts BRST
invariants $S(\lambda ,\theta)$ into supersymmetric and gauge-invariant components
$\langle S(\lambda ,\theta) \rangle$ \Berkovitsfe. Moreover, BRST-exact superfields
are annihilated, $\langle Q(E(\lambda ,\theta)) \rangle=0$ \Berkovitsfe. These
properties already motivate to study the BRST cohomology to foresee kinematic
factors in field-theory and string amplitudes in pure spinor superspace. From the
covariant BRST transformations of one-loop building blocks in \PPtwelve\ to
\PPfourteen,
\eqnn\PPeighteen
$$\eqalignno{
QM_{A}&= \sum_{XY=A} M_X M_Y\,,\cr
Q M_{A,B,C} &= \sum_{XY=A}\!\!(M_X M_{Y,B,C} - M_Y M_{X,B,C})
+ (A\leftrightarrow B,C)\,, &\PPeighteen \cr
QM^m_{A,B,C,D} &= \sum_{XY=A} (M_{X} M^m_{Y,B,C,D} - M_{Y} M^m_{X,B,C,D})
+k^m_{A} M_A M_{B,C,D} +(A \leftrightarrow B,C,D) \,,
}$$
one can recursively construct BRST-invariant scalars \EOMBBs\ such as
\eqnn\scalarBRST
$$\eqalignno{
C_{1|23,4,5} &\equiv M_1 M_{23,4,5} + M_{12}M_{3,4,5} - M_{13}M_{2,4,5}\,, \cr
C_{1|234,5,6} &\equiv M_1 M_{234,5,6} + M_{12}M_{34,5,6} + M_{123}M_{4,5,6} - M_{124}M_{3,5,6}\cr
&{}- M_{14}M_{23,5,6} - M_{142}M_{3,5,6} + M_{143}M_{2,5,6}\,, &\scalarBRST \cr
C_{1|23,45,6}
&\equiv M_1 M_{23,45,6} + M_{12}M_{45,3,6} - M_{13}M_{45,2,6} + M_{14}M_{23,5,6} - M_{15}M_{23,4,6}\cr
&{}+\big[ M_{124}M_{3,5,6} - M_{134}M_{2,5,6}+ M_{142}M_{3,5,6} - M_{143}M_{2,5,6}
- (4\leftrightarrow 5)\big]\,,\cr
}$$
and vectors \EOMBBs\ such as
\eqnn\PPfourtwo
$$\eqalignno{
C^m_{1|2,3,4,5} &\equiv M_1 M^m_{2,3,4,5} + \big[ k_2^m M_{12} M_{3,4,5}
+ (2\leftrightarrow 3,4,5) \big]\,, \cr
C^m_{1|23,4,5,6} &\equiv
M_1 M^m_{23,4,5,6} + M_{12} M^m_{3,4,5,6} - M_{13} M^m_{2,4,5,6} &\PPfourtwo\cr
&{} +\big[ k^m_3 M_{123}M_{4,5,6} + (3\leftrightarrow 4,5,6)\bigr]
- \big[ k^m_2 M_{132}M_{4,5,6} + (2\leftrightarrow 4,5,6)\bigr]\cr
&{}+\big[ k^m_4 M_{14}M_{23,5,6} + k^m_4 M_{142}M_{3,5,6} - k^m_4
M_{143}M_{2,5,6} + (4\leftrightarrow 5,6)\bigr]  \ .
}$$
Their gauge-invariant bosonic components $\langle C_{1|A,B,C} \rangle$ and $\langle
C^m_{1|A,B,C,D} \rangle$ determined from the zero-mode prescription \zerom\ can be
downloaded from \psweb. As detailed in section \sectwothree, the scalars in
\scalarBRST\ enter one-loop open-string amplitudes \Mafrakh\ but fail to explain
the hexagon anomaly in view of their BRST invariance $Q C_{1|A,B,C}=0$. The vectors
$C^m_{1|A,B,C,D}$ in turn are essential to efficiently represent the interactions
between left- and right-movers in closed-string amplitudes, see section \secfive.

\subsec BRST pseudo-invariants
\par\subseclab\reviewBRSTps

The hexagon gauge anomaly can be equivalently seen from a breakdown of BRST
invariance, see appendix \appgauge\ for further details. Hence, the superfields in
the anomaly kinematic factor \gaugevar\ 
\eqnn\Ydef
$$\eqalignno{
\cY_{A,B,C,D,E} &\equiv \half (\l\g^m \cW_A)(\l\g^n \cW_B)(\l\g^p
\cW_C)(\cW_D\g_{mnp}\cW_E) &\Ydef
}$$
are required to appear in the BRST variation of the six-point open-string amplitude. We will refer to 
gauge and BRST anomalies interchangeably in the rest of the paper.

The tensorial building block \PPfifteen\ selected by zero-mode arguments exhibits 
an anomalous BRST transformation of this type
in its trace component \cohom,
\eqnn\PPnineteen
$$\eqalignno{
Q M^{mn}_{A,B,C,D,E} &= \Big[  \sum_{XY=A} (M_{X} M^{mn}_{Y,B,C,D,E} 
- M_{Y} M^{mn}_{X,B,C,D,E}) &\PPnineteen \cr
&\quad\quad{} +2 k_{A}^{(m} M_{A} M^{n)}_{B,C,D,E} +(A \leftrightarrow B,C,D,E) \Big]  + \delta^{mn}\cY_{A,B,C,D,E}\ .
}$$
The same anomaly building block $\cY_{A,B,C,D,E} $ appears in the context
of a scalar anomaly current
whose single-particle version reads \cohom
\eqnn\PPfourseven
$$\eqalignno{
{\cal J}_{2|3,4,5,6} &\equiv \half A_2^m (M^m_{3,4,5,6} 
+ {\cal W}^m_{3,4,5,6})\,, &\PPfourseven\cr
Q{\cal J}_{2|3,4,5,6} &= k_2^m M_2 M^m_{3,4,5,6} + \big[ s_{23} M_{23} M_{4,5,6}
+ (3\leftrightarrow 4,5,6) \big]  + {\cal Y}_{2,3,4,5,6}\, .
}$$
While the above definition suffices for the six-point amplitude, a general
definition with multiparticle labels can be found in \cohom.

Instead of a BRST-invariant completion such as the scalars and vectors
in \scalarBRST\ and \PPfourtwo, the recursions of \cohom\ select the combinations
\eqnn\PPfourthree
\eqnn\PPone
$$\eqalignno{
C^{mn}_{1|2,3,4,5,6} &\equiv M_1 M^{mn}_{2,3,4,5,6} + 2 \big[ k_2^{(m} M_{12} M^{n)}_{3,4,5,6}
+ (3\leftrightarrow 4,5,6) \big]  &\PPfourthree\cr
& + 2 \big[ k_2^{(m} k_3^{n)} (M_{123}+M_{132}) M_{4,5,6} + (2,3|2,3,4,5,6) \big] \cr
P_{1|2|3,4,5,6} &\equiv M_1 {\cal J}_{2|3,4,5,6} + M_{12} k_2^m M^m_{3,4,5,6}
+ \big[ s_{23} M_{123} M_{4,5,6} + (3\leftrightarrow 4,5,6) \big] &\PPone
}$$
for the tensor \PPfifteen\ and the anomaly current in \PPfourseven. Since their BRST variations are exclusively 
furnished by the anomaly superfields in \Ydef,
\eqn\PPfiveone{
Q C^{mn}_{1|2,3,4,5,6} = -\d^{mn} V_1 {\cal Y}_{2,3,4,5,6} \ , \ \ \ \ \ Q P_{1|2|3,4,5,6}  = -  V_1 {\cal Y}_{2,3,4,5,6} \,,
}
these superfields are referred to as BRST {\it pseudo-invariants}. The motivation
for this terminology stems from the purely parity-odd bosonic components which
appear in the corresponding gauge variations such as \gaugevar\ \cohom. This ties
in with the linearized gauge transformations $e_1 \!\rightarrow \! k_1$ of the
expressions for $\langle P_{1|2|3,4,5,6} \rangle$ and $\langle C^{mn}_{1|2,3,4,5,6}
\rangle$ on the webpage~\psweb.

\subsec Worldsheet functions
\par\subseclab\secthreeone

\noindent String amplitudes augment kinematic factors with worldsheet integrals
where the former conspire to BRST invariants or pseudo-invariants once the integrals
are reduced to a basis. At one loop, the worldsheet integrand comprises
doubly-periodic functions of the insertion points $z_i$ of the vertex operators such as
the bosonic Green function on a genus-one surface with modular parameter $\tau$,
\eqn\GF{
G_{ij} \equiv G(z_{ij}|\tau) \equiv
\ln \left| { \theta_1(z_{ij}|\tau)\over \theta_1'(0|\tau)}
\right|^2 - {2\pi\over \tau_2}(\Im z_{ij})^2  \ ,
}
where $z_{ij} \equiv z_i -z_j$ and $\tau_2 \equiv \Im(\tau)$. Derivatives w.r.t.\ the first argument of $\theta_1(z|\tau)$ are 
interchangeably denoted by a tick and by $\partial \equiv \frac{\partial}{\partial z}$.
Exponentials of \GF\ give rise to the
Koba--Nielsen factor from the plane-wave correlator
seen for instance in \deltaAmptwo:
\eqn\KN{
{\cal I}(s_{ij}) \equiv \Big \langle \prod_{j=1}^{n} e^{i k_j \cdot x(z_j,\bar z_j)} \Big \rangle_{\tau} =
\prod_{i<j}^n \cases{ \exp \big[ \half\alpha' s_{ij} G_{ij} \big] &:
\ \hbox{closed string} \cr
\exp \big[ 2\ap  s_{ij} G_{ij} \big]&: \ \hbox{open string}
} \ ,
}
see \multmand\ for the conventions for Mandelstam invariants $s_{ij}$. As a main
result of this paper, we give a representation for the six-point open-string
integrand such that its BRST variation builds up the modular derivative of \KN\
required by the anomalous gauge variation \deltaAmp. For this purpose, we recall a
set of doubly-periodic functions $f^{(n)}(z_{ij}|\tau)\equiv f^{(n)}_{ij}$ with
$n=0,1,2,\ldots$ described in \emzv\ which were identified as a convenient language
for one-loop superstring amplitudes. In particular, it turns out that
\eqnn\fone
\eqnn\defftwo
$$\eqalignno{
f^{(1)}_{ij} &\equiv \partial_i G(z_{ij}|\tau) = 
 \partial \ln \theta_1(z_{ij}|\tau) +  2\pi i \, { \Im (z_{ij}) \over \tau_2}
&\fone\cr
f^{(2)}_{ij} &\equiv
{1\over 2} \Big\{  \Big( \partial \ln \theta_1(z_{ij}|\tau) +  2\pi i \, { \Im (z_{ij}) \over \tau_2} \Big)^2 - \wp(z_{ij}|\tau) \Big\}&\defftwo
}$$
with symmetries $f^{(1)}_{ij} = - f^{(1)}_{ji} $, $f^{(2)}_{ij} = f^{(2)}_{ji} $ and Weierstra\ss\ function \AWeyl
\eqn\Weier{
\wp(z|\tau) \equiv - \partial^2 \ln \theta_1(z|\tau) + {\theta_1'''(0|\tau) \over 3\theta_1'(0|\tau)}
}
suffice to describe the six-point amplitude. They are related via Fay's identity as \emzv,
\eqn\fones{
f^{(1)}_{ij}f^{(1)}_{ik} +
f^{(1)}_{ji}f^{(1)}_{jk} + f^{(1)}_{ki}f^{(1)}_{kj} = f^{(2)}_{ij} +f^{(2)}_{jk} +
f^{(2)}_{ki} \ ,
}
and one can show that the short-distance singularities of $(\partial \ln \theta_1)^2$ and $\wp$ 
drop out from \defftwo, rendering $f^{(2)}_{ij}$ non-singular as $z_{ij}  \rightarrow 0$. The relation of 
$f^{(2)}_{ij}$ with the $\tau$ derivative of the Green function \GF\ is
explained and applied in section \secfourone. The net result
\eqn\pretKN{
{\partial \over \partial t'} {\cal I}(s_{ij}) \sim {\cal I}(s_{ij}) \sum_{i<j}^6
s_{ij} f_{ij}^{(2)}
}
with $t' \equiv 1/t$ connects the derivative in moduli space
appearing in the gauge variation \deltaAmptwo\ with the function
$f^{(2)}_{ij}$ in the anomalous six-point correlator \PPthreeOne.

\newsec{The complete six-point amplitude of the open string}

In applying the pure spinor one-loop prescription \nmpsAmp,
the non-zero modes of the b-ghost
lead to cumbersome CFT calculations. One way to address this difficulty is to use
the BRST invariance of the pure spinor formalism as a guiding principle to write down
the answers directly. This will be done in this section for
the open-string six-point amplitude; the result contains
two classes of kinematic factors: BRST invariants ($\cK^C$) and
pseudo-invariants ($\cK^P$). Recalling the zero-mode prescription $\langle \ldots \rangle$ in
\zerom, our conventions are 
\eqn\fullKsix{
{\cal A}_6  =  \sum_{\rm top} G_{\rm top}  \langle A_6^{\rm top} \rangle \,,
\quad A_6^{\rm top}
\equiv \int_{0}^{\infty} \frac{\dd t}{t^5} \  \int_{\Delta_{\rm top}}\!\!\!
\dd z_2 \, \dd z_3 \, \ldots \, \dd z_6
\, {\cal I}(s_{ij})  \, ( {\cal K}_{6} ^C + {\cal K}_{6} ^P ) \ .
}
A separate analysis will be performed for each sector, and the pseudo-invariants 
$\cK^P$ will shortly be defined such as to make contact with the kinematic 
factor \gaugevar\ of the anomalous gauge variation.

\subsec The non-anomalous part of the worldsheet correlator
\par\subseclab\sectwothree

\noindent
A gauge-invariant subsector of one-loop open-string amplitudes which describes
the low-energy behavior
has been analyzed to all multiplicities in \Mafrakh. Its kinematic factors are captured by the BRST-closed scalars
$C_{i|A,B,C}$ in pure spinor superspace as exemplified in \scalarBRST.
Their derivation considers
only the zero-mode contributions from the b-ghost leading to the scalar
building blocks \PPtwelve\ and
follows from integration by parts identities of the worldsheet functions
associated to OPE singularities to reduce the integrals to a basis. More
specifically, products of worldsheet propagators \GF\ and $s_{ij}$ in \multmand,
\eqn\Xdef{
X_{ij} \equiv s_{ij} f^{(1)}_{ij} = s_{ij}\partial G_{ij}\,,
}
can be conveniently manipulated by discarding\foot{Boundary terms in $z_i$ do
not contribute since the exponential of $\ap s_{ij} G_{ij}$ vanishes as
$z_{ij}^{\ap s_{ij}}$ for $z_i \rightarrow z_j$. This is obvious if $s_{ij}$ has a
positive real part, whereas the vanishing for generic momenta follows from analytic
continuation \Richardsjg.} total derivatives acting on the Koba--Nielsen factor \KN:
\eqn\IBP{
\partial_{p} {\cal I}(s_{ij}) = \ap {\cal I}(s_{ij}) \sum_{q\neq p}
X_{pq} \ .
}
A basis of worldsheet functions in open- and closed-string correlators can be attained by
removing any explicit appearance of the
fixed insertion point $z_1$ along with $X_{1j}$ through the addition of total
derivatives \IBP\ with respect to $z_j$.

In terms of the worldsheet functions \Xdef\ and the BRST invariants
$C_{i|A,B,C}$, a permutation-invariant kinematic factor
for the six-point amplitude \fullKsix\ is given by \Mafrakh
\eqn\Ksix{
\eqalign{
{\cal K}_{6}^C &= \big[ X_{23} (X_{24}+ X_{34}) C_{1|234,5,6} + X_{24} (X_{23}+X_{43}) C_{1|243,5,6} 
+ (2,3,4 | 2,3,4,5,6) \big]  \cr
& +\big[ X_{23} X_{45} C_{1|23,45,6}+ X_{24} X_{35} C_{1|24,35,6}
+ X_{25} X_{34} C_{1|25,34,6} + (6 \leftrightarrow 5,4,3,2) \big]\,.
}}
As initially observed in \Mafrakh, the scalar BRST invariants $C_{i|A,B,C}$ can be re-expressed
in terms of color-ordered SYM tree amplitudes. At six points, the identities \EOMBBs
\eqnn\Csixa
\eqnn\Csixb
$$\eqalignno{
\langle C_{1|234,5,6} \rangle &= s_{56} \big[s_{45} A^{\rm YM}(1,2,3,4,5,6)
       -s_{35} A^{\rm YM}(1,2,4,3,5,6) \cr
       &-  s_{35} A^{\rm YM}(1,4,2,3,5,6)
       + s_{25} A^{\rm YM}(1,4,3,2,5,6)\big]&\Csixa \cr
\langle C_{1|23,45,6} \rangle &=   s_{46} s_{36} A^{\rm YM}(1,2,3,6,4,5)
       - s_{56} s_{36} A^{\rm YM}(1,2,3,6,5,4)  \cr
&  - s_{46} s_{26} A^{\rm YM}(1,3,2,6,4,5)
       + s_{56} s_{26} A^{\rm YM}(1,3,2,6,5,4)\,, &\Csixb\cr
}$$
allow to straightforwardly express all the polarization dependence of \Ksix\ in terms of $A^{\rm YM}(\ldots)$.
However, the above BRST-invariant integrand ${\cal K}_{6}^C$ cannot be the complete
answer for the six-point open string amplitude since it would imply manifest
gauge invariance\foot{We are grateful to Michael Green for insisting on a clarification of this point.}.
In the following, we will show how
the anomalous part of the amplitude can be described using the
BRST pseudo-invariants derived in \cohom\ and reviewed in section \reviewBRSTps.

\subsec{The anomalous part of the worldsheet correlator}

In order to correctly describe the anomalous part of
one-loop amplitudes, the kinematic factor ${\cal K}_6^P$ in \fullKsix\ cannot be BRST invariant.
According to \deltaAmp, its BRST variation must add up to a total
derivative in moduli space and reflect a parity-odd gauge variation. For this purpose, the notion of a {\it pseudo}
BRST cohomology was introduced in \cohom\ along with recursive method to construct
pseudo-invariants of arbitrary multiplicity and tensor rank.
Its scalar six-point representative $P_{1|2|3,4,5,6}$ has been defined in \PPone, and
its BRST variation $ - V_1 {\cal Y}_{2,3,4,5,6}$ in terms of the superfields \Ydef\ tie in with 
anomaly kinematic factor \gaugevar. That is why this superfield is suitable to describe the anomalous
gauge variation of the six-point integrand.

Using the above pseudo-invariants,
the anomalous part of the six-point correlator \fullKsix\ will be argued to be
\eqn\PPthreeOne{
{\cal K}_{6}^P = \big[ s_{12}f^{(2)}_{12} P_{1|2|3,4,5,6} + (2\leftrightarrow 3,4,5,6) \big]
+ \big[ s_{23}f^{(2)}_{23} P_{1|(23)|4,5,6} + (2,3|2,3,4,5,6) \big] \ ,
}
with $f^{(2)}_{ij}$ in \defftwo\ and
\eqn\Psym{
P_{1|(23)|4,5,6} \equiv P_{2|3|1,4,5,6}  - P_{2|1|3,4,5,6} + P_{1|2|3,4,5,6}\,.
}
Its BRST and gauge variations
\eqnn\PPthreetwo
$$\eqalignno{
Q{\cal K}_{6}^P &= - V_1 {\cal Y}_{2,3,4,5,6} \sum_{i<j}^6 s_{ij} f^{(2)}_{ij} \ , \ \ \ \ \ \
\delta {\cal K}_6^P  = - \Omega_1 {\cal Y}_{2,3,4,5,6}
\sum_{i<j}^6 s_{ij} f^{(2)}_{ij} +Q( \ldots)
&\PPthreetwo
}$$
will be identified as a boundary term in moduli space in section~\secfourone.
Therefore the anomaly cancellation for gauge group $SO(32)$ can be proven as in the
RNS formalism and will not be repeated here \refs{\Greenqs,\Cai,\Clavellifj,\Kubota}.

In contrast to the BRST-invariant kinematic factors $C_{i|A,B,C}$ in \Csixa\ and
\Csixb, the pseudo-invariant $P_{1|2|3,4,5,6}$ cannot be expressed in terms of
SYM tree-level subamplitudes. Two classes of tensor structures in its bosonic
components \psweb\ pose an obstruction:
\medskip
\item{1.} terms of the schematic form $(e_i \cdot k_j)^6$ where all the
six gluon polarization vectors $e_i$ with $i=1,2,\ldots,6$ are contracted
with an external momentum
\item{2.} parity-odd terms involving the ten-dimensional Levi-Civita tensor $\epsilon_{m_1 m_2 \ldots m_{10}}$
\medskip
\noindent It is easy to see from Feynman rules and worldsheet
supersymmetry that parity-even
contractions $(e_i \cdot k_j)^6$ are absent in tree amplitudes of both SYM and the
open superstring\foot{For an exploitation of this property in the RNS
formalism, see \refs{\Barreirodpa, \Barreiroaw}.}.

\subsubsec{Motivating the BRST pseudo-invariant worldsheet correlator}

The pseudo-invariants $P_{1|2|3,4,5,6}$ in \PPone\ are symmetric under
permutations of $3,4,5,6$ whereas the ``reference leg'' 1 is singled out by the choice of
unintegrated vertex $V_1$ in the amplitude prescription \nmpsAmp. This reasoning motivates to
associate $P_{1|2|3,4,5,6}$ with the worldsheet function $f^{(2)}_{12}$ in
\defftwo. Upon permutations in the integrated legs $2,3,\ldots,6$, this assigns natural 
kinematic companions $P_{1|2|3,4,5,6},\ldots,P_{1|6|2,3,4,5}$ to five instances
$f^{(2)}_{12},f^{(2)}_{13},\ldots ,f^{(2)}_{16}$ out of the 15 functions
$\{ f^{(2)}_{ij}, \ 1\leq i<j\leq 6\}$.

The form of the remaining kinematic factors can be inferred from the symmetry properties
of the anomalous correlator ${\cal K}_6^P$. In contrast to the permutation-invariant
expression for ${\cal K}_6^C$ in \Ksix, symmetry of ${\cal K}_6^P$ under
exchange of the unintegrated leg $(1\leftrightarrow 2)$ is slightly broken by the
anomaly. This can be traced back to the different response of unintegrated and integrated vertex operator
to gauge variations, see \gaugeVar. The anomalous BRST variation \PPfiveone\ makes reference to $V_1$ in the
prescription, and different choices of the unintegrated vertex are related by \cohom
\eqnn\PPtwentyone
$$\eqalignno{
Q {\cal Y}_{12,3,4,5,6} &= V_1 {\cal Y}_{2,3,4,5,6} - V_{2} {\cal Y}_{1,3,4,5,6} 
&\PPtwentyone
}$$
with a two-particle version ${\cal Y}_{12,3,4,5,6}$ of the anomaly building block \Ydef. This BRST variation
reproduces the antisymmetric part of the anomalous gauge variation \gaugevar, and a detailed account on 
the emergence of ${\cal Y}_{12,3,4,5,6}$ under antisymmetrization in $(1\leftrightarrow 2)$ can be found 
in appendix \appX. Indeed, the kinematic coefficient of the function $f^{(2)}_{12}=f^{(2)}_{21}$ is symmetric 
up to the BRST generator in \PPtwentyone\ \cohom,
\eqn\PPtwentytwo{
\langle P_{1|2|3,4,5,6} \rangle = \langle P_{2|1|3,4,5,6} -   {\cal Y}_{12,3,4,5,6} \rangle \ .
}
We interpret the superfield ${\cal Y}_{12,3,4,5,6}$ as an anomaly-transporting
term between external legs 1 and 2. Just as the anomalous gauge variation
\gaugevar, its bosonic
components are parity odd,
\eqn\PPtwentyfive{
\langle {\cal Y}_{12,3,4,5,6} \rangle =
 -\epsilon_{p_3 p_4 p_5 p_6 q_1 q_2 \ldots q_6} k_{3}^{p_3}
k_{4}^{p_4}k_{5}^{p_5}k_{6}^{p_6} e_1^{q_1} e_2^{q_2}  \cdots e_6^{q_6} \ ,
}
see appendix B of \cohom\ for a general argument.

Accordingly, the coefficient of $f^{(2)}_{23}$ cannot follow from a naive
relabeling of the legs in the combination $f^{(2)}_{12} \leftrightarrow
s_{12}P_{1|2|3,4,5,6}$ since $QP_{2|3|1,4,5,6}= -V_2 {\cal Y}_{1,3,4,5,6}$. However, we see from
\PPtwentyone\ that the anomalous BRST variation can be corrected via
${\cal Y}_{12,3,4,5,6}$. In view of \PPtwentytwo, the natural
candidate to multiply the function $f^{(2)}_{23}$ is $P_{1|(23)|4,5,6}$ in \Psym\ with
\eqnn\PPtwosix
$$\eqalignno{
QP_{1|(23)|4,5,6} &= -V_1 {\cal Y}_{2,3,4,5,6}\,. &\PPtwosix
}$$
The $2\leftrightarrow 3$ symmetry suggested by $f^{(2)}_{23}=f^{(2)}_{32}$ can be
checked to hold,
\eqnn\PPtwofour
$$\eqalignno{
\langle P_{1|(23)|4,5,6} - P_{1|(32)|4,5,6} \rangle&= \langle P_{1|2|3,4,5,6}-P_{2|1|3,4,5,6}
+ {\rm cyc}(1,2,3)\rangle &\PPtwofour \cr
&= -\langle {\cal Y}_{12,3,4,5,6}
+ {\cal Y}_{23,1,4,5,6} + {\cal Y}_{31,2,4,5,6} \rangle = 0 \  ,
}$$
where the cyclic combination of ${\cal Y}$'s in the second line is BRST trivial
under six-particle momentum conservation $k^m_{123456}=0$ \cohom\ 
(cf.\ \PPtwentyfive\ for the vanishing of the bosonic components). In the
interpretation of $\langle {\cal Y}_{12,3,4,5,6} \rangle$ as an anomaly
transportation term, the vanishing of \PPtwofour\ can be made plausible since the
second line describes an anomaly transportation around a closed loop $1 \rightarrow
2 \rightarrow 3 \rightarrow 1$. Note that an alternative cohomology representation of $P_{1|(23)|4,5,6}$ is
given by \cohom
\eqnn\alttwothree
$$\eqalignno{
\langle P_{1|(23)|4,5,6} \rangle &=
 \frac{1}{2}\big \langle (k_3^m - k_2^m)C^{m}_{1|23,4,5,6}  + P_{1|3|2,4,5,6} + P_{1|2|3,4,5,6} \cr
 & \ \ \ \ \ \ +
 \big[ s_{34} C_{1|234,5,6} + s_{24} C_{1|324,5,6}+(4\leftrightarrow 5,6) \big] \big
 \rangle \ .
 &\alttwothree
}$$
The symmetry properties of the anomalous correlator can be summarized as
\eqn\PPthreethree{
\langle {\cal K}_{6}^P \big|_{1\leftrightarrow2} - {\cal K}_{6}^P \rangle  =
\langle {\cal Y}_{12,3,4,5,6} \rangle \sum_{i<j}^6 s_{ij} f^{(2)}_{ij},
\qquad \langle {\cal K}_{6}^P \big|_{2\leftrightarrow3} - {\cal K}_{6}^P \rangle  = 0\ , 
}
see appendix \appX\ for a derivation from the amplitude prescription \nmpsAmp. 
The analysis in section \secfourone\ will also identify the failure of
permutation invariance in $\langle {\cal K}_{6}^P \rangle$ as a boundary~term.

In addition to the above plausibility arguments in superspace, we have explicitly
tested the anomalous correlator \PPthreeOne\ for consistency with the RNS
computation of the six-gluon amplitude. The technical aspects of this consistency
check are explained in appendix~\appA. The RNS computation must be carried out
separately for the parity-even and the parity-odd sector. The former is presented
in \appA.1, mostly guided by the results of \refs{\Tsuchiyava, \Stiebergerwk, \emzv}. The
parity-odd counterpart presented in appendix~\appA.2 largely follows the
computations in \Clavellifj\ apart from the presentation of worldsheet functions.
In the pure spinor representation of the correlator in \PPthreeOne, both
parity sectors are unified through the component expansion of the pseudo-invariants
$\langle P_{1|2|3,4,5,6} \rangle$ and $\langle P_{1|(23)|4,5,6} \rangle$.

Moreover, we have checked that the field-theory limit of the above six-point amplitude
reproduces the one-loop integrand of ten-dimensional SYM which has been derived 
in \symanom\ from cohomology arguments. Upon dimensional reduction to $D=4$, the
pseudoinvariant $P_{1|2|3,4,5,6}$ and therefore the entire anomalous correlator \PPthreeOne\
vanishes for MHV helicity configurations. Hence, the non-anomalous contribution \Ksix\ is
sufficient to derive the BCJ representation of MHV amplitudes in \HeWGF\
from the field-theory limit.

\subsec The BRST and gauge transformations as boundary terms
\par\subseclab\secfourone

In this subsection, we discuss the scalar integrals accompanying the anomalous BRST
and gauge variations \PPthreetwo\ of the six-point amplitude. In
particular, they are now demonstrated to describe boundary terms in the
moduli space of open-string worldsheets.

In order to relate the BRST variation \PPthreetwo\ of ${\cal K}_6^P$ to a total
derivative with respect to the modular parameter, it is worthwhile to express the
functions $f^{(2)}_{pq}$ in terms of the $\tau$ derivative of the bosonic Green
function \GF. For generic complex arguments,
the heat equation $4\pi i {\partial \theta_1(z|\tau) \over \partial \tau} =
\partial^2 \theta_1(z|\tau)$ obeyed by the theta function in \Weier\ implies that
\eqn\fversustau{
f^{(2)}_{pq} \equiv f^{(2)}(z_{pq}| \tau) = 2\pi i\Big(
{\partial G_{pq} \over \partial \tau} + {\Im z_{pq}\over \tau_2} \partial G_{pq} \Big)
+{\theta_1'''(0|\tau) \over 3\theta_1'(0|\tau)} - { \pi\over \tau_2} \ .
}
In a convenient parametrization of open-string worldsheets, the arguments $z_{pq},\tau$ of
the Green function \GF\ have constant real parts and are integrated
over their imaginary parts $\nu_{pq}\equiv \nu_p - \nu_q$ and $t$:
\eqn\parametr{
(z_{pq},\tau) \rightarrow 
\cases{ (i\nu_{pq},it) &:\ \hbox{$p$ and $q$ on the same cylinder boundary}  \cr
 (i\nu_{pq}+\half,it) &:\ \hbox{$p$ and $q$ on different cylinder boundaries}  \cr
  (i\nu_{pq},it+\half) &:\ \hbox{M\"obius strip}  \cr
}}
The integration domains $\Delta_{\rm top}$ for vertex insertions in \nmpsAmp\ and \fullKsix\ are then given by
\eqnn\Deltatop
$$\eqalignno{
\Delta_{P} &= \{  0 \leq  \nu_1 \leq \nu_2 \leq \ldots \leq \nu_{6} \leq t \}
\cr
\Delta_{N} &= \{  0 \leq  \nu_1 \leq \nu_2 \leq \ldots \leq \nu_{6} \leq 2t \}
&\Deltatop
\cr
\Delta_{NP} &= \{  0 \leq  \nu_1, \nu_2 \leq t \ {\rm and} \ 0 \leq  \nu_3 \ldots \leq \nu_{6} \leq t \} \ ,
}$$
where $\Delta_{P} $ and $\Delta_{N} $ are adapted to the single-traces over 
$t^{a_1}t^{a_2} t^{a_3}t^{a_4}t^{a_5}t^{a_6}$, and
$\Delta_{NP}$ refers to the non-planar cylinder diagram with color factor $G_{NP} = {\rm
tr}{(t^{a_1}t^{a_2})}{\rm tr}{(t^{a_3}t^{a_4}t^{a_5}t^{a_6})}$.
Hence, the functional dependence of $G_{pq}$ on the {\it real} parameters
$\nu_{pq}$ and $t$ is given as follows in the three inequivalent configurations:
\eqn\GFopen{
G_{pq} = G\big( i\nu_{pq}+ \delta | it + \vep  \big) \ ,
\ \ \ (\delta,\vep) = \cases{
(0,0) &:\ \hbox{$p$ and $q$ on the same cylinder boundary} \cr
(\half,0) &:\ \hbox{$p$ and $q$ on different cylinder boundaries} \cr
(0,\half) &:\ \hbox{M\"obius strip}
}}
Since the difference between planar and non-planar cylinders and the M\"obius strip
amounts to a constant shift of its arguments, $G_{pq}$ in \GFopen\ satisfies a universal
differential equation,
\eqn\PDE{
4\pi \Big( { \partial G_{pq}\over \partial t}
+ { \nu_{pq} \over t} {\partial G_{pq} \over \partial \nu_p} \Big)=
- \Bigl(   {\partial G_{pq} \over \partial \nu_p} \Bigr)^{\mkern-4mu 2}
- { \partial^2 G_{pq} \over \partial \nu_p^2} + c(t) \ .
}
On the right hand side, the definition \defftwo\ of $f^{(2)}_{pq}$ has been rewritten in terms 
of $\nu$-derivatives of $G_{pq}$.
The function $c(t)$ in \PDE\ does not depend on $\nu_p$ and will therefore drop out from the
later discussion. The differential operator on the left hand side can be recognized
as a derivative\foot{The partial derivative w.r.t. $t'$ is understood to be
evaluated at constant $\nu'$.} in the Jacobi transformed modular parameter:
\eqn\Jac{
t' \equiv {1\over t} \ , \ \ \ \nu' \equiv {\nu\over t}
\ \ \ \Rightarrow \ \ \ {\partial\over \partial t}
+ {\nu_{pq}\over t}   {\partial  \over \partial \nu_p} = -(t')^2{\partial \over \partial t'}\ .
}
The original modular parameter $t$ can be interpreted as the circumference of the
cylinder or the worldline length in the field-theory limit\foot{A pure spinor
description of the six-point one-loop amplitude in ten-dimensional SYM including
its hexagon anomaly can be found in \symanom.}. Its Jacobi transform $t'$, on the
other hand, describes the length of the cylinder or the proper time in the closed-string 
channel. Analogous statements hold for the M\"obius strip.

From \PDE\ and \Jac, one can derive a universal relation for
$f^{(2)}_{pq}$ analogous to \fversustau, 
\eqn\fopen{
f^{(2)} \big( i\nu_{pq}+ \delta | it + \vep  \big)=  -2\pi (t')^2{\partial \over \partial t'}G\big( i\nu_{pq}+ \delta | it + \vep  \big) + \frac{c(t)}{2} \ .
}
This allows to rewrite the $t'$ derivative of the Koba--Nielsen factor \KN\ in
terms of $f^{(2)}_{ij}$,
\eqn\tKN{
{\partial \over \partial t'}{\cal I}(s_{ij}) =
-{\ap \over 2\pi (t')^2}{\cal I}(s_{ij})
\sum_{i<j}^6 s_{ij} f^{(2)}_{ij} \ ,
}
which is valid for all topologies and where $c(t)$ in \fopen\ cancels by momentum conservation $\sum_{i<j}^6
s_{ij}=0$. Moreover, the pattern of Mandelstam variables and $f^{(2)}_{ij}$ on the
right hand side reproduces the anomalous BRST transformation \PPthreetwo\ of the six
point correlator,
\eqn\ttKN{
Q \big( {\cal I}(s_{ij}) \, {\cal K}_6 \big) 
= V_1 {\cal Y}_{2,3,4,5,6} { 2\pi (t')^2  \over \ap}
{\partial \over \partial t'}{\cal I}(s_{ij}) \ .
}
Together with the Jacobi transformed integration measure $\dd t = - { \dd t' \over
(t')^2}$, one can finally identify the BRST anomaly of the six point
amplitude in \fullKsix\ as a boundary term in $t'$:
\eqnn\Qalltop
$$\eqalignno{
Q A_6^{\rm top} &= { 2\pi  \over \ap}V_1 {\cal Y}_{2,3,4,5,6} 
\int_{0}^{\infty} \dd t' {\partial \over \partial t'}
\int_{\Delta_{\rm top}} \dd z'_2 \, \dd z'_3 \, \ldots \, \dd z'_6 \, {\cal I}(s_{ij}) \ ,
&\Qalltop
}$$
where the transformation $\dd z_j = it \dd z_j'$ has compensated for the factor of $t^{-5}$ in \KN. 
Note that modular invariance of the Koba--Nielsen factor allows to collectively replace 
$G(i\nu_{pq} | it )\rightarrow G(\nu_{pq}'| it' )$. By the universality of \fopen, this analysis 
is valid for all topologies of open-string worldsheets
and the anomaly is canceled for the gauge group $SO(32)$ \Greenqs.

\newsec The complete six-point amplitude of the closed string
\par\seclab\secfive

\noindent This section is devoted to the six-point one-loop amplitude among massless 
closed-string states of type IIA/IIB superstring theories.
Before presenting the six-point function we revisit the five-point amplitude 
result of \Greenbza\ to rewrite its kinematics in a factorized form.

\subsec{The one-loop five point function for closed strings}

In \Greenbza\ the pure spinor representation of
the five-point closed-string amplitude in both type IIA/IIB was obtained\foot{The RNS
and GS representations can be found in \BjerrumBohrvc\ and \Richardsjg, respectively.}
(in the type IIA the chirality of the right-movers is reversed)
\eqnn\cloneTmp
$$\eqalignno{
{\cal M}_5 &= \int { \dd^2 \tau \over \tau_2^5} \int \dd^2 z_2 \ldots \dd^2 z_5
\ {\cal I}(s_{ij}) \, \big(
{\cal K}_5 \tilde {\cal K}_5 + {\pi \over \tau_2} {\cal L}_5\big)\,, &\cloneTmp
}$$
where ${\cal K}_5$ is the open-string five point correlator
and ${\cal L}_5$ encodes the interactions between the left- and right-movers
(marked with tilde),
\eqnn\cltwo
\eqnn\clone
$$\eqalignno{
{\cal K}_5 &= X_{23} C_{1|23,4,5} + (2,3|2,3,4,5), &\cltwo\cr
{\cal L}_5 & =
M_1 M^m_{2,3,4,5} \tilde M_1 \tilde M^m_{2,3,4,5}
 + \big[ s_{12} M_{12} M_{3,4,5} \tilde M_{12} \tilde M_{3,4,5}
 + (2\leftrightarrow 3,4,5) \big]  \cr
&  +  \big[ s_{23} M_{1} M_{23,4,5} \tilde M_{1} \tilde M_{23,4,5}
- s_{23} C_{1|23,4,5} \tilde C_{1|23,4,5}
+ (2,3|2,3,4,5) \big]\,,
&\clone
}$$
see \PPtwelve\ and \PPfourteen\
for the definitions of $M_{A,B,C},M^m_{A,B,C,D}$ and \scalarBRST\ for
$C_{1|A,B,C}$ with
\eqn\Cfivep{
\langle C_{1|23,4,5} \rangle = s_{45}\big[s_{24} A^{\rm YM}(1,3,2,4,5)
- s_{34}A^{\rm YM}(1,2,3,4,5) \big]\,.
}
The characteristic coefficient
${\pi\over \tau_2}$ signals the mixing between left- and right-movers and arises
from either the contraction $\Pi^m(z)\bar\Pi^n(\bar z)$ or from left-moving derivatives
acting on right-moving propagators in integration by parts identities,
\eqn\clfour{
\Pi^m(z_i) \bar \Pi^n(\bar z_j) \rightarrow \delta^{mn} {\pi \over \tau_2},\qquad
\p_i \bar f^{(1)}_{ij} = - {\pi\over \tau_2}\,.
}
While the amplitude \cloneTmp\ is BRST invariant
the kinematic factor ${\cal L}_5$ is not manifestly BRST closed.
However, by adding terms to ${\cal L}_5$ that vanish in the cohomology
one arrives at a manifestly BRST invariant expression (the vector $C^m_{1|2,3,4,5}$ is
reviewed\foot{The shorthand $C^m_{1|2,3,4,5} \tilde C^m_{1|2,3,4,5}$ was
assigned a different meaning in \Greenbza\ and differs from the
right-hand side of \factfive\ by
$s_{23} C_{1|23,4,5} \tilde C_{1|23,4,5} + (2,3|2,3,4,5)$.}
in section~\reviewBRST),
\eqn\factfive{
{\cal L}_5 + \Big[QD_{1|2|3,4,5}\tilde M_{12}\tilde M_{3,4,5}
+ M_{12} M_{3,4,5} \tilde Q\tilde D_{1|2|3,4,5} + (2\leftrightarrow 3,4,5)\Big]
= C^m_{1|2,3,4,5} \tilde C^m_{1|2,3,4,5},
}
where (note that $\langle {\cal Y}_{1,2,3,4,5}\rangle = 0$ in the five-particle momentum phase space) \cohom,
\eqnn\cleight
$$\eqalignno{
D_{1|2|3,4,5} &\equiv {\cal J}_{2|1,3,4,5} + k_2^m M_{12,3,4,5}^m
 + \big[ s_{23} M_{123,4,5} + (3\leftrightarrow 4,5) \big], &\cleight\cr
QD_{1|2|3,4,5} &= {\cal Y}_{1,2,3,4,5} + k_2^m M_1 M_{2,3,4,5}^m - s_{12} M_{12}M_{3,4,5}
+ \big[ M_1 M_{23,4,5} + (3\leftrightarrow 4,5)\big].
}$$
Therefore the five-point amplitude \clone\ becomes
\eqn\clseven{
{\cal M}_5  = \int { \dd^2 \tau \over \tau_2^5}
\int \dd^2 z_2 \,\ldots \, \dd^2 z_5 \ {\cal I}(s_{ij}) \,
\Big\{ {\cal K}_5 \tilde {\cal K}_5
+ {\pi \over \tau_2} C^m_{1|2,3,4,5} \tilde C^m_{1|2,3,4,5} \Big\} \, ,
}
up to the $Q$-exact terms in \factfive\ that do not contribute upon zero-mode integration.
 This representation is manifestly BRST invariant (since $QC_{1|A,B,C} = Q
C^m_{1|A,B,C,D}=0$) and organizes the kinematic dependence in a factorized form
w.r.t.\ left- and right-movers.

The compactness and manifest BRST invariance of \clseven\ demonstrate the virtue
of vectorial BRST invariants to describe closed-string amplitudes. From the
five-point example, one can anticipate that BRST (pseudo-)invariants of rank $r$
find a natural appearance in closed-string amplitudes at higher multiplicity $r+4$, along
with $r$ powers of ${\pi \over \tau_2}$. In the subsequent, this expectation is
confirmed for the six-point amplitude.

\subsec{The six-point closed-string correlator}

The six-point closed-string correlator ${\cal M}_6$ combines the doubling of its open-string
counterpart ${\cal K}_6={\cal K}^C_6+{\cal K}^P_6$ with an extended set of left-right
interactions,
\eqn\clten{
{\cal M}_6 =\int { \dd^2 \tau  \over \tau_2^5}
\int \dd^2 z_2 \,\ldots \, \dd^2 z_6 \ {\cal I}(s_{ij}) \,
\Big\{ {\cal K}_6 \tilde {\cal K}_6 + {\pi \over \tau_2}
{\cal K}^m_6 \tilde {\cal K}^m_6 + \Big({\pi \over \tau_2}\Big)^2 {\cal L}_6 \Big\}\,,
}
where (see \Ksix\ and \PPthreeOne\ for the expressions of $\cK_6^C$ and $\cK_6^P$)
\eqnn\cloneone
$$\eqalignno{
\cK_6  & ={\cal K}_6^C+{\cal K}_6^P &\cloneone\cr
\cK_6^m &= X_{23} C^m_{1|23,4,5,6} + (2,3|2,3,4,5,6) \ .
}$$
Note that ${\cal K}_6^m$ resembles the five-point open string correlator
\cltwo\ where the scalar invariants $C_{1|23,4,5}$ are replaced by their vector
counterparts \PPfourtwo. The appearance
of ${\cal K}^m_6 \tilde {\cal K}^m_6$ has been carefully checked
by keeping track of all the sources of ${\pi\over\tau_2}$ shown in \clfour.

Finally, the kinematic factor ${\cal L}_6$ along with the quadratic piece ${\pi^2
\over \tau^2_2}$ in \clten\ contains the two-tensor generalization of left-right
contractions supplemented by a quadratic expression of the pseudo-invariants \PPone,
\eqn\clonetwo{
{\cal L}_6 = \half C^{mn}_{1|2,3,4,5,6} \tilde C^{mn}_{1|2,3,4,5,6}
- \big[ P_{1|2|3,4,5,6} \tilde P_{1|2|3,4,5,6} + (2\leftrightarrow 3,4,5,6) \big] \ .
}
The pseudo-invariants in \clonetwo\ obstruct a representation
of ${\cal L}_6$ as a tensor contraction of
the form ${\cal K}_6^{mn} \tilde {\cal K}_6^{mn}$ but their presence compensates
the anomalous BRST transformation \PPfiveone\ of the tensor $C^{mn}_{1|2,3,4,5,6}$
such that $Q{\cal L}_6 = 0$. In addition,
we will show in the next section that the form of \clonetwo\
is fixed by the low-energy limit.

Since the functions $f^{(n)}$ and $\tau_2^{-1}$ have modular weight $(n,0)$ \emzv\ and $(1,1)$,
respectively, $\cK_6$ and $\cK_6^m$ carry modular weight $(2,0)$ and $(1,0)$ such that the 
expression in \clten\ manifests modular invariance of the closed-string amplitude.

\subsec{Low-energy limits and S-duality of type IIB}

In this subsection, we discuss the low-energy limit of one-loop amplitudes among
massless closed-string states and relate it to the S-duality implications
in type IIB theory.

As explained in \refs{\Greenpv, \Greenuj, \Richardsjg, \Greenbza}, the momentum
dependence of torus integrals of the form \clseven\ and \clten\ can be
split into analytic and non-analytic contributions\foot{The interplay
of the analytic and non-analytic parts of the amplitude as well as subtle
ambiguities at higher $\ap$ order and their resolution are discussed in 
\Greenuj.}. The leading analytic behavior $\alpha'\rightarrow 0$ follows
unambiguously by setting ${\cal I}(s_{ij}) \rightarrow 1$ after taking
the kinematic poles due to integration over $\dd^2 z {\cal I}(s_{ij})
\partial G_{23} \bar \partial G_{23} \rightarrow 2\pi \dd r_{23}\,r_{23}^{\ap s_{23}-1}$
with $r_{23} \equiv |z_{23}|$ into account, see e.g. \refs{\Richardsjg, \Greenbza}.
This gives rise to low-energy limits
\eqnn\clonefive
\eqnn\clonesix
\eqnn\cloneseven
$$\eqalignno{
{\cal M}_4 \big|_{\alpha'\rightarrow 0} &=|C_{1|2,3,4} |^2 &\clonefive \cr
{\cal M}_5 \big|_{\alpha'\rightarrow 0} &=  \big[ s_{23} |C_{1|23,4,5} |^2 +
(2,3|2,3,4,5) \big] +  |C^m_{1|2,3,4,5} |^2 &\clonesix\cr
{\cal M}_6 \big|_{\alpha'\rightarrow 0} &=
 \big[ s_{23} |C^m_{1|23,4,5,6} |^2  + (2,3|2,3,4,5,6) \big]
+ {\cal L}_6 &\cloneseven
}$$
\vskip-32pt
$$
+ \big[ s_{23} s_{45} |C_{1|23,45,6} |^2 + s_{24} s_{35} |C_{1|24,35,6} |^2
+ s_{25} s_{34} |C_{1|25,34,6} |^2 + (6\leftrightarrow 5,4,3,2) \big]
$$
\vskip-32pt
$$
+ \big[ s_{23} s_{34} |C_{1|234,5,6} |^2 +s_{24} s_{43} |C_{1|243,5,6} |^2
+s_{23} s_{24} |C_{1|324,5,6} |^2 + (2,3,4|2,3,4,5,6) \big]
$$
with the shorthand notation $|C^m_{1|23,4,5,6} |^2\equiv C^m_{1|23,4,5,6} \tilde C^m_{1|23,4,5,6} $ and
obvious generalizations. In the six-point case, the expressions \Ksix\ and \cloneone\ for ${\cal
K}_6^C$ and ${\cal K}_6^m$ have been inserted into \clten, whereas the
expression \clonetwo\ for ${\cal L}_6$ is treated as unknown at this point and
will be derived in the subsequent. Note that the anomalous part ${\cal K}_6^P$ of
the open-string correlator does not contribute to the low-energy limit due to the
non-singular nature of the $f^{(2)}_{ij}$ as $z_i \rightarrow z_j$, see the
discussion below \defftwo.

At four and five points, the
type IIB graviton components of \clonefive\ and \clonesix\ are proportional to the $\ap^3$
order of tree-level amplitudes \refs{\Greensw,\Greenbza}. They originate from the $R^4$ operator in the type IIB low-energy 
effective action whose tensor structure is determined by supersymmetry and
whose coefficient (determined by S-duality) is given by the non-holomorphic Eisenstein
series $E_{3/2}$ \refs{\Greentv, \Greenby, \Sinhazr}.
The non-linear extension of $R^4$ equally affects multiparticle amplitudes at the
$\ap^3$ order at tree level and in the low-energy limit at one loop and leads to
the following S-duality prediction,
\eqnn\cloneeight
$$\eqalignno{
\langle {\cal M}_N\big|^{\rm IIB}_{\alpha'\rightarrow 0} \rangle &=
c_q\!\!\!\!\sum_{\sigma ,\rho \in S_{N-3}} A^{\rm YM}(1,\sigma(2,3,\ldots,N-2),N,N-1) &\cloneeight\cr
&\quad \times  (S_0 M_3)_{\sigma,\rho}  \tilde A^{\rm YM}(1,\rho(2,3,\ldots,N-2),N-1,N)
}$$
whose proportionality constant $c_q$ does not depend on the multiplicity $N$. The
right-hand side borrows the notation of \Schlottererny\ for the low-energy
expansion of tree-level amplitudes involving $N$ closed-string states. The entries
of the $(N-3)! \times (N-3)!$ matrices $S_0$ and $M_3$ are
polynomials of degree $N-3$ and $3$ in the dimensionless Mandelstam invariants $\ap
(k_{i}\cdot k_{j})$, and $S_0$ is the momentum kernel
\BjerrumBohrhn\ which appears in the field-theory limit of the KLT formula
\refs{\Kawaixq, \Bernsv}. The matrix $M_3$ captures the $\ap^3$ order in
the low-energy expansion of genus-zero worldsheet integrals\foot{The explicit form
of these matrices for
multiplicity $N \leq 7$ and the building blocks for $N=8,9$ can be downloaded from \WWW. 
Initially addressed via hypergeometric functions \refs{\Oprisawu, \Stiebergerbh},
the $\alpha'$-corrections at tree level for any 
multiplicity can be recursively generated from the Drinfeld associator \Broedelaza. 
The organization of these integrals in $(N-3)! \times (N-3)!$ matrices has been essential to reveal the structure 
of the $\alpha'$-expansion \Schlottererny, see \Drummondvz\ for relations to the
associator.}.

A slightly modified argument applies to the components of \cloneeight\ which violate
the $U(1)$ R-symmetry of type IIB supergravity. This is indicated by the subscript
$q$ of the proportionality constant $c_q$ in \cloneeight. The simplest non-vanishing
amplitude with $U(1)$ violation occurs at multiplicity five and charge $q=\pm 2$,
involving for instance four gravitons and one axio-dilaton, see \refs{\Boelszr,
\Boelsjua, \Greenbza} for its $\ap$-expansion. It was argued via S-duality and
confirmed through explicit calculation that the constants in \cloneeight\ for
charges $q=0,\pm2$ are related by $c_{\pm 2}=-{1 \over 3} c_0$ \Greenbza. An analogous
discussion of the low-energy limit of two-loop five-point amplitudes and their dependence on 
R charges can be found in \GomezUHA.

Since the coefficient ${\cal L}_6$ of ${\pi^2 \over \tau_2^2}$ in the six-graviton
amplitude \clten\ contributes to the low-energy limit \cloneseven, it can be
determined from the S-duality prediction \cloneeight. More precisely, the form of
${\cal L}_6$ in \clonetwo\ is inferred from the following reasoning.

The double contraction $\Pi^m \Pi^n \bar \Pi^p \bar \Pi^q \rightarrow 2
\delta^{m(p} \delta^{q)n}\big({\pi \over \tau_2})^2 $ gives rise to a contribution
of the form
$\half M_1 M^{mn}_{2,3,4,5,6}\tilde M_1 \tilde M^{mn}_{2,3,4,5,6}$
whose unique BRST pseudo-invariant completion is given by
${\cal L}_6 = \half C^{mn}_{1|2,3,4,5,6}\tilde C^{mn}_{1|2,3,4,5,6}+\cdots$ \cohom.
However, the BRST variation \PPfiveone\ and
the trace relation $\d_{mn}\tilde C^{mn}_{1|2,3,4,5,6} = 2 \tilde P_{1|2|3,4,5,6} +
(2\leftrightarrow 3,4,5,6)$ \cohom\ yield
\eqn\clonenine{
Q C^{mn}_{1|2,3,4,5,6}\tilde C^{mn}_{1|2,3,4,5,6}=
- 2 V_1 {\cal Y}_{2,3,4,5,6} \big[ \tilde P_{1|2|3,4,5,6} + (2\leftrightarrow 3,4,5,6) \big].
}
Now the S-duality prediction relates the low-energy
limit of the closed-string amplitude and the tree-level $\ap^3$ terms via \cloneeight. Demanding the low-energy limit
to be BRST invariant and permutation symmetric\foot{Demonstrating permutation invariance of \cloneseven\ with ${\cal
L}_6$ given by \clonetwo\ requires the canonicalization techniques in section~11 of \cohom.}
uniquely fixes ${\cal L}_6$ to the form \clonetwo. A component evaluation
for six external gravitons confirms the matching with the tree-level amplitude at
order $\ap^3$.

\subsec{The BRST variation as a boundary term}
\par\subseclab\secfivethree

\noindent It will be demonstrated in this section\foot{We are grateful to Michael
Green for fruitful discussions which led to the results of this section.} that the
BRST (or gauge) variation of the closed-string amplitude \clten\
gives rise to a total derivative in moduli space.

BRST invariance of ${\cal K}_6^m$ and ${\cal L}_6$ implies that
\eqn\cltwenty{
Q{\cal M}_6 =- V_1{\cal Y}_{2,3,4,5,6}\int { \dd^2 \tau  \over \tau_2^5}
\int \dd^2 z_2 \,\ldots \, \dd^2 z_6 \ {\cal I}(s_{ij}) \,
\sum_{i<j}^6 s_{ij} f^{(2)}_{ij} \, \tilde {\cal K}_6  \ .
}
Using the representation \fversustau\ of the $f^{(2)}_{ij}$ function, the factor of
$\sum_{i<j}^6 s_{ij} f^{(2)}_{ij}$ can be expressed in terms of derivatives of the
Koba--Nielsen factor with respect to $\tau$ and $z_j$:
\eqnn\cltwothree
\eqnn\cltwotwo
$$\eqalignno{
{1 \over 2\pi i} {\cal I}(s_{ij})\sum_{i<j}^6 s_{ij} f^{(2)}_{ij} &=
{\cal I}(s_{ij}) \Big( \sum_{i<j}^6 s_{ij} {\partial \over \partial \tau} G_{ij}
+{1 \over \tau_2} \sum_{i=1}^6 \Im z_i \sum_{j \neq i}^6 s_{ij} \partial G_{ij} \Big)
&\cltwothree \cr
&= {2 \over \alpha'} \Big( {\partial \over \partial \tau}
+ \sum_{p=2}^6 { \Im z_{p1} \over \tau_2} \partial_p  \Big) {\cal I}(s_{ij}) \ .
&\cltwotwo
}$$
The second step is based on translation invariance $\partial_1 {\cal I} =
-\sum_{j=2}^6 \partial_j {\cal I}$. It turns out that the differential operator in
\cltwotwo\ annihilates the right-moving correlator $\tilde {\cal K}_6$ since \emzv
\eqn\barder{
  \partial \bar f^{(k)}_{ij} = - { \pi \over \tau_2}  \bar f^{(k-1)}_{ij} \ ,
  \ \ \ \ {\partial \bar f^{(k)}_{ij}\over \partial  \tau} =
  {\pi \Im z_{ij}\over \tau_2^2}   \bar f^{(k-1)}_{ij} \ , \ \ \ \  k=1,2
}
with $\bar f^{(0)}_{ij} \equiv 1$ imply that its constituents $\bar f^{(2)}_{ij}$ 
and $\bar f^{(1)}_{ij} \equiv \bar \partial G_{ij}$ satisfy
\eqn\cltwofour{
 \Big( {\partial \over \partial \tau}
 + \sum_{p=2}^6{ \Im z_{p1} \over \tau_2} \partial_p \Big) \bar f^{(k)}_{ij} = 0 \quad k=1,2 \ .
}
Hence, the BRST variation in \cltwenty\ can be rewritten as
\eqn\cltwosix{
Q{\cal M}_6 =- {4\pi i \over \ap} V_1 {\cal Y}_{2,3,4,5,6}\int { \dd^2 \tau  \over \tau_2^5}
\int \dd^2 z_2 \,\ldots \, \dd^2 z_6 \
\Big( {\partial \over \partial \tau}
+ \sum_{p=2}^6 { \Im z_{p1} \over \tau_2} \partial_p  \Big)
\big({\cal I}(s_{ij})\, \tilde {\cal K}_6 \big) \ .
}
In order to identify this as total derivatives, we have to commute the differential operators
${\partial \over \partial \tau} $ and $\partial_p$ past the factors of ${1 \over \tau_2^5}$ and
${ \Im z_{p1} \over \tau_2}$, respectively. The commutators 
\eqn\cltwoseven{
\Big[  {1 \over \tau_2^5}  , {\partial \over \partial \tau}  \Big] =
-{5i \over 2\tau_2^6} \ , \quad
\Big[ {\Im z_{p1}\over \tau_2^6}  , \partial_p  \Big] = {i \over 2\tau_2^6}
}
mutually cancel after summing $p$ over $2,3,\ldots,6$, so we conclude\foot{The action of 
${\partial \over \partial \tau}$ on the $\tau$ dependent integration domain for $z_j$ drops
out because the resulting boundary term $z_j = \tau$ is suppressed by the Koba--Nielsen factor.}
that the BRST variation of the six point function is a surface term in both
$\tau$ and $z_p$:
\eqnn\cltwonine
$$\eqalignno{
Q{\cal M}_6 &=- {4\pi i \over \ap} V_1 {\cal Y}_{2,3,4,5,6}\int
\dd^2 \tau \Bigl\{  {\partial \over \partial \tau}
{1 \over \tau_2^5}  \int \dd^2 z_2 \,\ldots \, \dd^2 z_6 \
{\cal I}(s_{ij})\, \tilde {\cal K}_6  \cr
&  \ \ \ \ \ 
+ \sum_{p=2}^6  \int \dd^2 z_2 \,\ldots \, \dd^2 z_6 \,
\p_p { \Im z_{p1} \over \tau_2^6 }  {\cal I}(s_{ij})\,
\tilde {\cal K}_6 \Bigr\}\,.&\cltwonine
}$$
The surface integral over the vertex insertions in the second line vanishes because
the torus has no boundaries while the vanishing of the surface integral on moduli space
follows from modular invariance \refs{\Hayashi,\Kutasov,\LercheNP}.


\newsec{Conclusion and outlook}

In this work we combined the one-loop cohomology analysis of \cohom\ with the
worldsheet functions studied in \emzv\ to write down the complete six-point
one-loop amplitudes of the open and closed string. In doing so, we supplemented
the BRST-invariant six-point correlator \Ksix\ that captures the worldsheet
singularities among the external vertices with the non-singular pseudo-invariant
correlator \PPthreeOne. The {\it pseudo} BRST invariance allows us to describe the
hexagon gauge anomaly in pure spinor superspace while the non-singular worldsheet
functions capture the regular parts of the correlator. Their composition given in
\PPthreeOne\ is such that its non-vanishing gauge variation \PPthreetwo\ gives rise
to a total derivative in moduli space (leading to the usual mechanism of anomaly
cancellation \refs{\Greenqs,\Cai}). This condition fixes the superspace form of the anomaly-containing
part of the open-string correlator \PPthreeOne\ and reproduces the bosonic results
from earlier analyses within the RNS framework \refs{\Tsuchiyava,\Clavellifj}. 

The (pseudo-)invariant vector and tensor building blocks from the open string allow for
elegant representations for closed-string one-loop amplitudes. As elaborated in section~4,
any basis integral of the closed string is accompanied by a manifestly (pseudo-)invariant
kinematic factor quadratic in the open-string (pseudo-)invariants. In order to arrive at the
novel six-point result, in addition to an OPE-driven derivation of the singular part of the 
correlator, we also used S-duality considerations to completely fix its regular terms. This 
organizing principle for closed-string one-loop amplitudes has a natural extension beyond
maximal supersymmetry, see \BBS\ for examples in orbifold compactifications.

While the results of this paper demonstrate the value of the (pseudo-)cohomology 
framework of \cohom, it is imperative to derive them from first
principles within the pure spinor formalism. This endeavor is expected to require a more
in-depth understanding of how the non-zero modes of the b-ghost contribute to the
final expressions in analogy to the RNS supercurrent in appendix B.2. 
These contributions are currently poorly understood and give
rise to difficulties in extending the results for higher-loop amplitudes in \refs{\GomezSLA, \GomezUHA} 
beyond their low-energy limits.

Furthermore, for gauge groups different than $SO(32)$, additional boundary terms along the
lines of \BerkovitsRPA\ arise from regularizing the divergent modular integral\foot{We are grateful 
to Michael Green for enlightening discussions on this point.}. These boundary terms
give rise to worry about additional BRST anomalies and ambiguities associated with the choice of 
the regulator ${\cal N}$ for the non-compact space of pure spinors \NMPS. Subtleties of this type 
are not addressed in this work, and their treatment in a manifestly supersymmetric formalism is left 
as an interesting open problem.


\bigskip \noindent{\bf Acknowledgements:} We are very grateful to Tim Adamo, Nathan Berkovits,
Eduardo Casali, Yu-tin Huang, David Skinner, Stephan Stieberger and in particular Michael Green 
for inspiring discussions. CRM wishes to acknowledge support from NSF grant number 
PHY 1314311 and the Paul Dirac Fund.
OS thanks Marcus Berg and Igor Buchberger for collaboration on related topics and
 the Institute of Advanced Studies in Princeton for kind hospitality during finalization
of this work. The research leading to these results has received funding from the
European Research Council under the European Community's Seventh Framework
Programme (FP7/2007-2013) / ERC grant agreement no. [247252]. Also, this material
is supported by the National Science Foundation under Grant No. PHY-1066293 and the
hospitality of the Aspen Center for Physics.

\appendix{A}{Permutation behavior of the open-string correlator}
\applab\appX

\noindent
In this appendix, we derive the asymmetry \PPthreethree\ of the six-point correlator ${\cal K}_6^P$ from the prescription \nmpsAmp\ for open-string amplitudes. It will be demonstrated that an exchange of the unintegrated vertex operator $V_1 U_2 \rightarrow V_2 U_1$ yields a boundary term accompanied by the kinematic factor $\langle {\cal Y}_{12,3,4,5,6} \rangle$ in \Ydef\ with parity-odd bosonic components \PPtwentyfive.

The key tool is the multiparticle version of the integrated vertex operator in \vertices,
\eqn\Uonetwo{
{\cal U}_{12} \equiv  \p\theta^\a {\cal A}^{12}_\a + \Pi^m {\cal A}^{12}_m + d_\a {\cal W}_{12}^\a + {1\over 2}N_{mn}{\cal F}_{12}^{mn} \ ,
}
where the linearized superfields in $U_j$ are promoted to their
multiparticle versions defined in section \secmultpar. As a consequence of
their multiparticle equations of motion, the BRST
variation $QU = \partial V$ generalizes to \EOMBBs
\eqn\QUonetwo{
Q{\cal U}_{12} = V_1 U_2 - V_2 U_1+ \partial M_{12} \ .
}
When inserting the left hand side into the amplitude prescription \nmpsAmp\ in the place of $V_1 U_2$, the analysis of $\{Q,b\}$ around \deltaAmp\ can be repeated to show that
\eqnn\AUonetwo
$$\eqalignno{
& \sum_{\rm top} G_{{\rm top}}\int_0^{\infty} \dd t
\int_{\Delta_{\rm top}}\!\!\!\!\!  \dd z_2 \, \dd z_3 \, \ldots \, \dd z_n \,\, \langle{ {\cal N} (b,\mu)
  \, (Q{\cal U}_{12})\,   \prod_{j=3}^n U_j (z_j)}\rangle  &\AUonetwo\cr
& \ \ \ \ =  -\sum_{\rm top} G_{\rm top}\int_0^{\infty} \dd t  {\partial \over \partial t} \int_{\Delta_{\rm top}} \dd z_2 \, \ldots \, \dd z_6 \,  \langle{ {\cal N}
   \, {\cal U}_{12}  \,\prod_{j=3}^6 U_j (z_j)}\rangle\,.
}$$
After integrating $d_{\alpha_1}d_{\alpha_2}d_{\alpha_3}d_{\alpha_4}d_{\alpha_5} \rightarrow (\lambda
\g^m)_{\alpha_1}(\lambda \g^n)_{\alpha_2}(\lambda \g^p)_{\alpha_3}(\g_{mnp})_{\alpha_4\alpha_5}$ \fiveptNMPS\ for the only term $(d{\cal W}_{12}) (d W_3)\ldots(d W_6)$ with a sufficient number of $d_\alpha$ zero modes, \AUonetwo\ evaluates to
\eqn\bdyU{
\langle {\cal Y}_{12,3,4,5,6} \rangle
\sum_{\rm top} G_{\rm top}
\int_0^{\infty}{ \dd t \over t^5}   \int_{\Delta_{\rm top}} \dd z_2 \, \ldots \, \dd z_6 \,  
\sum_{i<j}^6 s_{ij} f^{(2)}_{ij}
{\cal I}(s_{ij}) 
}
by the definition \Ydef\ of the anomaly superfield. Using the Jacobi-transformed variables $t'={1\over t}$ and $z' = {z\over t}$ in intermediate steps, the modular derivative of the Koba--Nielsen factor ${\cal I}(s_{ij})$ in \KN\ has been evaluated via \tKN\ and gives rise to the functions $f^{(2)}_{ij}$ defined in \defftwo. In the conventions of \fullKsix, one can read off the contribution of $\langle {\cal Y}_{12,3,4,5,6} \rangle \sum_{i<j}^6 s_{ij} f^{(2)}_{ij}$ to the antisymmetric part of the kinematic factor ${\cal K}_6$ from \bdyU.

This needs to be compared with the amplitude prescription \nmpsAmp\ involving the right hand side of \QUonetwo: The total derivative $\partial M_{12}$ decouples by the suppression of boundary terms in $z_j$ via $z_{ij}^{\alpha' s_{ij}}$, and the leftover term $V_1 U_2 - V_2 U_1$ yields the desired difference between ${\cal K}_6$ and its image under $(1\leftrightarrow 2)$. This completes the proof of \PPthreethree.

\appendix{B}{Comparison with the RNS computation}
\applab\appA

\noindent
We have checked the six-point open-string amplitude \PPthreeOne\ in pure spinor superspace to reproduce the 
gluon amplitude from the RNS formalism upon component expansion \psweb. Since this comparison rests on the availability
of both expressions in a basis of worldsheet integrals, we will sketch the underlying integral reduction on the RNS
side in this appendix.

\subsec{The parity-even part}

The RNS prescription for the parity-even part of one-loop amplitudes is given by
\eqnn\RNSeven
$$\eqalignno{
A_{6,{\rm even}}^{{\rm top}} \sim \int_{0}^{\infty}& \dd t \int_{\Delta_{\rm top}} \dd z_2 \, \dd z_3\, \ldots \, \dd z_6 \, \sum_{\nu =1,2,3}  (-1)^{\nu} \left( { \theta_{\nu+1}(0,\tau) \over \theta'_1(0,\tau) } \right)^4  \cr
& \times  \langle V_1(e_1,k_1,z_1) \, V_2(e_2,k_2,z_2) \, \ldots \, V_6(e_6,k_6,z_6) \rangle_{\nu,\tau} \ ,
&\RNSeven
}$$
where $V_i$ denotes the vertex operator of the gluon in the superghost picture zero:
\eqn\vzero{
V_1(e_1,k_1,z_1) \equiv e_1^m \big[ \partial x_m(z_1)  + 2\ap k_1^n \psi_n \psi_m(z_1) \big] e^{i k_1\cdot x(z_1)} \ .
}
The bracket $\langle \ldots \rangle_{\nu,\tau}$ instructs to evaluate the correlator in \RNSeven\
on a genus-one Riemann surface with modular parameter $\tau$, and $\nu=1,2,3$ encode the even spin
structures of the worldsheet spinors $\psi^m$ associated with partition functions $(-1)^{\nu} \left( {
\theta_{\nu+1}(0 | \tau) \over \theta'_1(0 |\tau) } \right)^4$ \Tsuchiyava.

Correlators among $x^m$ and $\psi^m$ can be straightforwardly evaluated using
Wick contractions $x^m(z_i) x^n(z_j) \rightarrow -2\ap \delta^{mn} G_{ij}$ and
$\psi^m(z_i) \psi^n(z_j) \rightarrow \delta^{mn} S_\nu(z_{ij}|\tau)$. The latter give
rise to spin structure dependent Szeg\"o kernels 
\eqn\szego{
S_\nu(z|\tau) \equiv { \t_1'(0|\tau) \t_{\nu+1}(z|\tau) \over  \t_{\nu+1}(0|\tau) \t_1(z|\tau)}
}
with $\nu=1,2,3$. Together with the partition function in the first line
of \RNSeven, the summation over spin structures can be described by the
following building block
\eqn\calG{
{\cal G}_n(x_1,x_2,\ldots,x_n|\tau) \equiv \sum_{\nu=1,2,3} (-1)^{\nu}
\left( { \t_{\nu+1}(0 | \tau) \over \t'_1(0 | \tau) } \right)^4
S_\nu(x_1|\tau) S_\nu(x_2|\tau) \ldots S_\nu(x_n|\tau)
}
with $x_1+x_2+\ldots+x_n=0$. As is well known, correlators with less than eight $\psi^m$ yield a
vanishing spin sum, and ${\cal G}_4$ is the first instance where Riemann identities yield a
non-vanishing result,
\eqn\vanis{
{\cal G}_{n\leq 3}(x_1,x_2,\ldots,x_n|\tau)=0 \ , \qquad {\cal G}_{4}(x_1,x_2,x_3,x_4|\tau)=1 \ ,
}
reflecting maximal spacetime supersymmetry. Representatives at multiplicity five and
higher have been evaluated in \refs{\Tsuchiyava, \Stiebergerwk} using Fay
trisecant identities in slightly different guises. The results of these references
are equivalent to \emzv\
\eqnn\GGfive
\eqnn\GGsix
$$\eqalignno{
 {\cal G}_{5}(x_1,x_2,x_3,x_4,x_5|\tau)&= \sum_{i=1}^5  f^{(1)}_i &\GGfive \cr
  {\cal G}_{6}(x_1,x_2,x_3,x_4,x_5,x_6|\tau)&= \sum_{i=1}^6 f^{(2)}_i + \sum_{i<j}^6 
 f^{(1)}_i f^{(1)}_j \ ,
   &\GGsix
}$$
where $f^{(k)}_{i} \equiv f^{(k)}(x_{i}|\tau)$ are defined by \fone\ and \defftwo, respectively.

In order to cast the RNS amplitude \RNSeven\ into the same basis of integrals as seen in the
expression \Ksix\ and \PPthreeOne\ for ${\cal K}_6^C$ and ${\cal K}_6^P$ in pure
spinor superspace, we organize the integral reduction into three steps:
\medskip
\item{(i)} {\bf elimination of double derivatives}:
Bilinears in $\partial x^m(z_i)$ from the vertex operator \vzero\ contract to
a double derivative $\partial^2 G_{ij}$ of the Green function \GF. Since
there is always a partner term $\ap s_{ij} (\partial G_{ij})^2$ with the same
tensor structure from the fermionic part $\sim \psi^2$ of the vertex operators,
the double pole of $\partial^2 G_{ij},(\partial G_{ij})^2 \sim {1 \over z_{ij}^2}$
turns out to be spurious. This can be seen from a total derivative relation involving
the Koba--Nielsen factor ${\cal I}$ from \KN:
\eqn\ztwoIBP{
\p_i \big( \p G_{ij} {\cal I} \big) = \big[ \p^2 G_{ij} + \ap s_{ij} (\p G_{ij})^2 + \ap \p G_{ij} \sum_{p\neq i,j} X_{ip} \big] {\cal I}
}
The residue of the double pole must be proportional to $ (1-\ap s_{ij})$ since it would otherwise signal tachyon propagation.
\item{(ii)} {\bf partial fraction relations}:
Step (i) and \GGsix\ leave two topologies of bilinears in the propagator: $\p G_{ij}\p G_{ik}$
with $j\neq k$ and an overlapping leg $i$ as well as the disconnected configuration
$\p G_{ij} \p G_{pq}$ with all of $i,j,p,q$ distinct. The former requires
an application of the Fay identity \fones\ before the pattern of functions $X_{ij} (X_{ik}+X_{jk})$
seen in \Ksix\ and suitable for step (iii) is manifest:
\eqn\pfloop{
\p G_{ij} \p G_{ik} = \frac{ s_{jk}(f^{(2)}_{ij}+f^{(2)}_{ik}+f^{(2)}_{jk}) }{s_{ijk}}
+ \frac{ X_{ij}(X_{ik}+X_{jk}) }{s_{ij} s_{ijk}} + \frac{ X_{ik}(X_{ij}+X_{kj}) }{s_{ik} s_{ijk}} \ .
}
\item{(iii)} {\bf integration by parts}:
As explained below \IBP, the minimal set of worldsheet functions $X_{ij}$ is obtained by
eliminating any instance of $X_{1j}$ by discarding derivatives of the Koba--Nielsen factor
\KN\ w.r.t. $z_j$. This amounts to two equivalent manipulations after step (ii):
\eqnn\IBPsixb
$$\eqalignno{
X_{12} (X_{13}+X_{23}) &=  (X_{23}+X_{24}+X_{25}+X_{26}) (X_{34}+X_{35}+X_{36})
+ \p_2(\ldots) + \p_3(\ldots)
  \cr
X_{12}X_{34} &= (X_{23}+X_{24}+X_{25}+X_{26}) X_{34} + \p_2(\ldots)
&\IBPsixb
}$$
\medskip

\noindent The $f^{(2)}_{ij}$ functions from \GGsix\ and step (ii) do not admit
further simplification, in particular the five instances of $f^{(2)}_{1j}$ cannot be reduced to $f^{(2)}_{pq}$ 
and $X_{pq}$ with $p,q\neq 1$.
After performing the above steps, the agreement of bosonic components in the two
formalisms,
\eqn\evtop{
A_{6,{\rm even}}^{{\rm top}} = \langle A^{{\rm top}}_6 \rangle
\big|_{{\rm parity-even}} \ ,
}
can be checked along with each instance of
$f^{(2)}_{ij}, X_{ij}(X_{ik}+X_{jk})$ and $X_{ij} X_{pq}$, 
see \fullKsix\ for the definition of the right hand side.

\subsec The parity-odd part
\par\subseclab\appAtwo

\noindent The parity-odd sector of the RNS six-point amplitude stems from
the spin structure of $\psi^m$ with anti-periodic boundary conditions
along both cycles of the Riemann surface. In this case, zero modes
of the $\beta,\g$ ghosts as well as the ten components of $\psi^m$ have
to be saturated in the path integral. This gives rise to the amplitude
prescription
\eqnn\RNSodd
$$\eqalignno{
A_{6,\te{odd}}^{\te{top}} \sim \frac{1}{2} \int_{0}^{\infty}& \dd t \int_{\Delta_{\rm top}} \dd z_2 \, \dd z_3\, \ldots \, \dd z_6 &\RNSodd \cr
& \times  \langle \p x_p(z_0) \psi^p(z_0) \widehat V_1(e_1,k_1,z_1) \, V_2(e_2,k_2,z_2) \, \ldots \, V_6(e_6,k_6,z_6) \rangle_{\tau} \ ,
}$$ 
where $\widehat V_1(e_1,k_1,z_1)$ denotes the gluon vertex operator
in the picture of superghost charge $-1$, and the zero mode integration 
for the $\beta,\g$ system has already been carried
out \Grosspd\ 
\eqn\vone{
\widehat V_1(e_1,k_1,z_1)  \equiv e_1^m   \psi_m(z_1)  e^{i k_1\cdot x(z_1)}\,.
}
The worldsheet supercurrent $\p x_p \psi^p$ is a remnant of a picture changing
operator whose position $z_0$ drops out from the correlator.
In the evaluation of the correlator \RNSodd, the Wick contraction
of $\psi^m$ is adjusted to the spin structure
$\psi^m(z_i) \psi^n(z_j) \rightarrow \delta^{mn} \p G_{ij}$, and the
zero mode integration amounts to absorbing
\eqn\zeroeps{
\psi^{m_1}(w_1) \psi^{m_2}(w_2) \ldots \psi^{m_{10}}(w_{10}) \rightarrow \epsilon^{m_1 m_2\ldots m_{10}} \ ,
}
independently on $w_j$. After simplifying the parity-odd kinematic factors
and eliminating the double derivatives of $G_{0j}$ in a way similar to \ztwoIBP,
\eqn\ztwoIBPP{
\p^2 G_{0j} {\cal I}(s_{ij})= \ap {\cal I}(s_{ij}) \p G_{0j} \sum_{p \neq j} X_{jp}
+ \p_j(\ldots) \ ,
}
we arrive at the following expression for \RNSodd:
\eqn\RNSoddtwo{
A_{6,\te{odd}}^{\te{top}} = \int_{0}^{\infty} \frac{\dd t}{t^5} \int_{\Delta_{\rm top}} \dd z_2 \,  \ldots \, \dd z_6 \,  {\cal I}(s_{ij})
\,\Big\{ \sum_{2\leq p< q}^6 {\cal E}_{pq} \big[ \eta_{0pq} - \eta_{01p} - \eta_{01q} - (\p G_{01})^2\big] \Big\}.
}
The worldsheet functions contained in $\eta_{ijk} \equiv \p  G_{ij} \p  G_{ik}
+ {\rm cyc}(i,j,k)$ can be rewritten as $f^{(2)}_{ij}+f^{(2)}_{jk}+f^{(2)}_{ki}$
via \fones, and the shorthand ${\cal E}_{pq}$ encoding the polarization
dependence is defined by (permutations of)
\eqnn\epsone
$$\eqalignno{
{\cal E}_{23} &\equiv \frac{1}{2}e_1^m \big[ (e_2\cdot k_3) k_2^n - s_{23} e_2^n \big]
\epsilon_{mnr_3 s_3\ldots r_6 s_6} k_3^{r_3} e_3^{s_3} \ldots k_6^{r_6} e_6^{s_6}
+ (2\leftrightarrow 3) \ .
&\epsone
}$$
The sum of the ten inequivalent ${\cal E}_{pq}$ in \RNSoddtwo\ can be written as an
antisymmetrization in eleven vector indices such that
\eqn\epstwo{
\sum_{2\leq p<q}  {\cal E}_{pq} = 0\,.
}
This is crucial to cancel the contributions of $\p G_{01}^2$ and
$f^{(2)}_{01}$ in $\eta_{0pq} - \eta_{01p} - \eta_{01q} - (\p G_{01})^2$
such that the position of the supercurrent in \RNSodd\ drops out. We are left with
\eqnn\RNSoddthree
$$\eqalignno{
A_{6,\te{odd}}^{\te{top}} = \int_{0}^{\infty}&\frac{ \dd t}{t^5} \int_{\Delta_{\rm top}} \dd z_2 \, \dd z_3\, \ldots \, \dd z_6 \,  {\cal I}(s_{ij})
\,\Big\{ \sum_{2\leq p< q}^6  f^{(2)}_{pq}  {\cal E}_{pq}
+ \sum_{p=2}^6 f^{(2)}_{1p} {\cal E}_{p}  \Big\}
 \ ,
&\RNSoddthree
}$$
where the functions $f^{(2)}_{1p}$ pick up polarization dependencies such as
\eqn\epsthree{
 {\cal E}_{2} \equiv - \sum_{q=3}^6 {\cal E}_{2q} = \frac{1}{2} \big[ (e_1\cdot k_2 )k_2^m e_2^n+ (e_2\cdot k_1) k_1^m e_1^n - s_{12}e_1^m e_2^n \big]
  \epsilon_{mn r_3 s_3 \ldots r_6 s_6} k_3^{r_3} e_3^{s_3} \ldots k_6^{r_6} e_6^{s_6}
}
along with $f^{(2)}_{12}$. The kinematic factors \epsone\ and \epsthree\ are easily seen to be gauge
invariant w.r.t. $e_j^m \rightarrow k_j^m$ for $j\neq1$. However,
the variation $e_1^m \rightarrow k_1^m$ in the first leg
(represented by the vertex operator \vone\ of superghost picture $-1$) gives rise to
\eqnn\epsfour
\eqnn\epsfive
$$\eqalignno{
 {\cal E}_{pq} \big|_{e_1^m \rightarrow k_1^m} &= s_{pq}  \times \epsilon_{m_2 n_2 \ldots m_6 n_6} k_2^{m_2} e_2^{n_2} \ldots k_6^{m_6} e_6^{n_6}
 &\epsfour
 \cr
 {\cal E}_{p} \big|_{e_1^m \rightarrow k_1^m} &= s_{1p} \times  \epsilon_{m_2 n_2 \ldots m_6 n_6} k_2^{m_2} e_2^{n_2} \ldots k_6^{m_6} e_6^{n_6} \ ,
 &\epsfive
}$$
since an $\epsilon$ contraction of all the six momenta $k_1,k_2,\ldots,k_6$ vanishes
by momentum conservation. The resulting gauge anomaly
\eqnn\epssix
$$\eqalignno{
A_{6,\te{odd}}^{\te{top}}  \big|_{e_1^m \rightarrow k_1^m} &= \epsilon_{m_2 n_2 \ldots m_6 n_6} k_2^{m_2} e_2^{n_2} \ldots k_6^{m_6} e_6^{n_6}  \cr
& \ \ \ \ \ \ \times  \int_{0}^{\infty} \frac{\dd t}{t^5} \int_{\Delta_{\rm top}} \dd z_2 \, \dd z_3\, \ldots \, \dd z_6 \,  {\cal I}(s_{ij})
\,  \sum_{1\leq p< q}^6 s_{pq} f^{(2)}_{pq}
&\epssix
}$$
is the fingerprint of the anomalous BRST variation \PPthreetwo\ on the bosonic
components, see appendix~\appgauge\ for a superspace discussion of gauge
variations. The parity-odd part \RNSoddthree\ of the RNS amplitude agrees with the
bosonic components of the superamplitude in \fullKsix,
\eqn\compar{
A_{6,\te{odd}}^{\te{top}} =   \langle A^{\te{top}}_6 \rangle \big|_{\te{parity-odd}} \ ,
}
which is found by comparing the coefficient of any $f^{(2)}_{ij}$.

\appendix{C}{Gauge transformation versus BRST transformation}
\applab\appgauge

\noindent In this appendix, it is demonstrated that linearized gauge transformations of the external 
states are encoded in the BRST variations of the kinematic factors. We thereby 
prove the equivalence of the anomalous BRST and gauge variations \PPthreetwo\ of the six-point open-string 
amplitude \Ksix\ and \PPthreeOne.

\subsec Gauge variation of multiparticle superfields

The response of linearized SYM superfields to a superspace gauge transformation $\delta_i$ in
particle $i$ is given by
\eqn\gaugeone{
\delta_i A^i_\alpha = D_\a \Omega_i , \ \ \ \ \ \delta_i A^i_m = k^i_m \Omega_i , \ \ \ \ \ \delta_i W_i^\alpha= \delta_i F_i^{mn} = 0 \,,
}
for scalar superfields $\Omega_i$, leading to the variations \gaugeVar\ of the 
massless vertex operators. For the choice $\Omega_i= e^{ik_i\cdot x}$, 
the gauge transformation \gaugeone\ amounts to  a
transverse gluon polarization $e_i^m \rightarrow k_i^m$. 

The recursive construction of multiparticle superfields ${\cal A}^P_\alpha, {\cal
A}_P^m,{\cal W}_P^\alpha,{\cal F}^{mn}_P$ in \BGdef\ to \PPten\ determines
their linearized gauge variation from \gaugeone. As pioneered in appendix~B
of \cohom\ and generalized in \LeeUPY, multiparticle gauge transformations
are conveniently captured by multiparticle gauge scalars
\eqn\gauA{
{\cal G}_P \equiv {1\over s_P} \sum_{XY=P} \big[ {\cal G}_Y (k_Y \cdot {\cal A}_X) - {\cal G}_X (k_X \cdot {\cal A}_Y) \big] \ .
}
Performing a linearized gauge transformation \gaugeone\ in a single external
leg (say leg $i=1$) amounts to the initial condition
${\cal G}_j =  \delta_{1,j}\Omega_1$ for the recursion \gauA, i.e.\ to 
having only one non-vanishing single-particle scalar ${\cal G}_j$. The
induced gauge transformation of multiparticle superfields is given by \LeeUPY
\eqnn\gauB
\eqnn\gauC
\eqnn\gauD
\eqnn\gauE
$$\eqalignno{
\delta_{{\cal G}} {\cal A}_\alpha^P &= D_\alpha {\cal G}_P + \sum_{XY=P} ({\cal G}_X {\cal A}_\alpha^Y - {\cal G}_Y {\cal A}_\alpha^X) 
&\gauB \cr
\delta_{{\cal G}}  {\cal A}_P^m &=  k^m_P {\cal G}_P + \sum_{XY=P} ({\cal G}_X {\cal A}^m_Y - {\cal G}_Y {\cal A}^m_X) 
&\gauC \cr
\delta_{{\cal G}}  {\cal W}^\alpha_P &=  \sum_{XY=P} ({\cal G}_X {\cal W}^\alpha_Y - {\cal G}_Y {\cal W}^\alpha_X) 
&\gauD \cr
\delta_{{\cal G}}  {\cal F}_P^{mn} &=  \sum_{XY=P} ({\cal G}_X {\cal F}_Y^{mn} - {\cal G}_Y {\cal F}^{mn}_X)  \ ,
&\gauE
}$$
where the $\delta_{{\cal G}}$ operation reduces to the linearized variations \gaugeone\ in 
any leg $i$ for appropriate choices of initial conditions. As a consequence, the one-loop 
building blocks in \McalVdef\ to \PPfifteen\ transform as
\eqnn\gauF
\eqnn\gauG
\eqnn\gauH
\eqnn\gauI
$$\eqalignno{
\delta_{{\cal G}}  M_A &= Q{\cal G}_A + \sum_{XY=A} ({\cal G}_X M_Y - {\cal G}_Y M_X) &\gauF \cr
\delta_{{\cal G}}  M_{A,B,C} &=  \sum_{XY=A} ({\cal G}_X M_{Y,B,C} - {\cal G}_Y M_{X,B,C}) + (A\leftrightarrow B,C) &\gauG \cr
\delta_{{\cal G}}  M^m_{A,B,C,D} &=  \sum_{XY=A} ({\cal G}_X M^m_{Y,B,C,D} - {\cal G}_Y M^m_{X,B,C,D}) \cr
&\ \ \ \ + k_A^m {\cal G}_A M_{B,C,D} + (A\leftrightarrow B,C,D) &\gauH \cr
\delta_{{\cal G}}  M^{mn}_{A,B,C,D,E} &=  \sum_{XY=A} ({\cal G}_X M^{mn}_{Y,B,C,D,E} - {\cal G}_Y M^{mn}_{X,B,C,D,E}) \cr
& \ \ \ \ + 2k_A^{(m} {\cal G}_A M^{m)}_{B,C,D,E} + (A\leftrightarrow B,C,D,E)  \ ,&\gauI 
}$$
and the anomaly current \PPfourseven\ exhibits the following gauge variation:
\eqn\gauJ{
\delta_{{\cal G}}  J_{2|3,4,5,6} = k^m_2 {\cal G}_2 M^m_{3,4,5,6} + \big[ s_{23} {\cal G}_{23} M_{4,5,6} + (3\leftrightarrow 4,5,6) \big] \ .
}
The multiparticle response to gauge variations can be conveniently interpreted by assembling both the 
gauge scalars ${\cal G}_P$ and the multiparticle superfields ${\cal A}_\alpha^P,{\cal A}_m^P,\ldots$ in a 
generating series: While latter solve the non-linear equations of motion of ten-dimensional SYM \MafraGIA, 
the resummation of the gauge scalars encodes their non-linear gauge transformations. The recursion \gauA\ is obtained
by demanding the non-linear gauge transformations to preserve the Lorentz-gauge condition for the 
generating series of ${\cal A}_m^P$ \LeeUPY. The benefits of certain different choices of multiparticle 
gauge scalars are discussed in the reference.

\subsec Gauge variation of BRST (pseudo-)invariants

For all of the kinematic building blocks $\{ M_A, M_{A,B,C}, \ldots\}$ in the amplitudes under discussion, the BRST variations \PPeighteen, \PPnineteen\ and \PPfourseven\ closely resemble the gauge variations \gauF\ to \gauJ. It is therefore not surprising that BRST invariants such as the scalars $C_{1|A,B,C}$ and the vectors $C^m_{1|A,B,C,D}$ in section \reviewBRST\ give rise to a $Q$-exact gauge variation
\eqn\gauK{
\delta_{{\cal G}}  C_{1|A,B,C} = Q\big[  C_{1|A,B,C}  \big|_{M_P \rightarrow {\cal G}_P} \big] \ , \ \ \ \ 
\delta_{{\cal G}}  C^m_{1|A,B,C,D} = Q\big[  C^m_{1|A,B,C,D}  \big|_{M_P \rightarrow {\cal G}_P} \big] \ ,
}
leading to vanishing components,
\eqn\gauKLM{
\langle \delta_{{\cal G}}  C_{1|A,B,C} \rangle =0 \ , \ \ \ \ 
\langle \delta_{{\cal G}}  C^m_{1|A,B,C,D} \rangle = 0 \ .
}
For instance the five-point BRST invariant $C_{1|23,4,5} = M_1 M_{23,4,5} + M_{12} M_{3,4,5} - M_{13} M_{2,4,5}$ translates into the gauge variation $\delta_{{\cal G}}  C_{1|23,4,5} = Q({\cal G}_1 M_{23,4,5} + {\cal G}_{12} M_{3,4,5} - {\cal G}_{13} M_{2,4,5})$ captured by the replacement $M_P \rightarrow {\cal G}_P$.

For the tensor $M^{mn}_{A,B,C,D,E}$ and the anomaly current ${\cal J}_{2|3,4,5,6}$, however, the superfields ${\cal Y}_{A,B,C,D,E}$ in their BRST variations \PPnineteen\ and \PPfourseven\ do not have any correspondent in the gauge variations \gauI\ and \gauJ. That is why the gauge transformation of their pseudo-invariant completions $C^{mn}_{1|2,3,4,5,6}$ and $P_{1|2|3,4,5,6}$ in \PPfourthree\ and \PPone\ exhibit anomalous admixtures
\eqnn\gauL
\eqnn\gauM
$$\eqalignno{
\delta_{{\cal G}}  C^{mn}_{1|2,3,4,5,6}  &= Q\big[  C^{mn}_{1|2,3,4,5,6}  \big|_{M_P \rightarrow {\cal G}_P} \big] - \delta^{mn} {\cal G}_1 {\cal Y}_{2,3,4,5,6}
&\gauL \cr
\delta_{{\cal G}}  P_{1|2|3,4,5,6}  &= Q\big[  P_{1|2|3,4,5,6}  \big|_{M_P \rightarrow {\cal G}_P} \big] - {\cal G}_1 {\cal Y}_{2,3,4,5,6} \ .
&\gauM
}$$
The components $\langle \delta_{{\cal G}}  C^{mn}_{1|2,3,4,5,6}\rangle$ and $\langle \delta_{{\cal G}} P_{1|2|3,4,5,6} \rangle$ only depend on ${\cal G}_1$ whereas any other ${\cal G}_{j\neq 1}$ drops out. This reflects the initial observation in section \sectwotwo\ that gauge transformations of the integrated vertices $U_2,\ldots,U_6$ annihilate the six-point amplitude while the unintegrated vertex $V_1$ yields the anomaly \deltaAmptwo\ upon variation to $Q\Omega_1$. Setting ${\cal G}_j \rightarrow \delta_{j,1} \Omega_1 e^{ik_jx}$ reproduces the anomaly kinematic factor $K$ in \gaugevar:
\eqn\gauNNN{
\langle \delta_{{\cal G}}  C^{mn}_{1|2,3,4,5,6}\rangle \rightarrow  -\frac{1}{2} \delta^{mn} K
\ , \ \ \ \ \langle \delta_{{\cal G}}  P_{1|2|3,4,5,6} \rangle \rightarrow  -\frac{1}{2}  K
\ .
}
The gauge anomaly in \gauL\ and \gauM\ obviously matches the anomalous BRST variations
\PPfiveone\ upon adjusting ${\cal G}_j \rightarrow V_j$ in the non-exact part. Hence, the mechanism of anomaly 
cancellation is completely analogous for gauge and BRST transformations, see \Greenqs\ for open
strings, and section~\secfivethree\ for the closed-string discussion.

\listrefs
\bye